\documentclass[twocolumn,trackchanges]{aastex6}
\usepackage{multirow}
\usepackage{enumitem}
\usepackage{amsmath}
\usepackage{threeparttable} 
\hypersetup{linkcolor=blue,citecolor=blue,filecolor=blue,urlcolor=blue}
\usepackage{url}
\def\gapp{\ifmmode\stackrel{>}{_{\sim}}\else$\stackrel{>}{_{\sim}}$\fi}

\usepackage[bookmarks=false]{hyperref}

\begin{document}
\title{The implementation of a Fast-Folding pipeline for long-period pulsar searching in the PALFA survey}
\author{
E.~Parent\altaffilmark{1},
V.~M.~Kaspi\altaffilmark{1},
S.~M.~Ransom\altaffilmark{2},
M.~Krasteva\altaffilmark{3},
C.~Patel\altaffilmark{1},
P.~Scholz\altaffilmark{4},
A.~Brazier\altaffilmark{5,6},
M.~A.~McLaughlin\altaffilmark{7,8},
M.~Boyce\altaffilmark{1},
W.W.~Zhu\altaffilmark{9,10},
Z.~Pleunis\altaffilmark{1},
B.~Allen\altaffilmark{11,12,22},
S.~Bogdanov\altaffilmark{13}, 
K.~Caballero\altaffilmark{14},
F.~Camilo\altaffilmark{15}, 
R.~Camuccio\altaffilmark{14},
S.~Chatterjee\altaffilmark{5}, 
J. M.~Cordes\altaffilmark{5},
F.~Crawford\altaffilmark{16}, 
J. S.~Deneva\altaffilmark{17}, 
R.~Ferdman\altaffilmark{18}, 
P. C. C.~Freire\altaffilmark{10},
J. W. T.~Hessels\altaffilmark{19,20},
F. A.~Jenet\altaffilmark{21},
B.~Knispel\altaffilmark{11,12}, 
P.~Lazarus\altaffilmark{10}, 
J.~van Leeuwen\altaffilmark{19,20}, 
A. G.~Lyne\altaffilmark{23}, 
R.~Lynch\altaffilmark{2}, 
A.~Seymour\altaffilmark{10}, 
X.~Siemens\altaffilmark{22}, 
I. H.~Stairs\altaffilmark{24},
K.~Stovall\altaffilmark{25},
J.~Swiggum\altaffilmark{22}, 
}

\altaffiltext{1}{Dept.~of Physics and McGill Space Institute, McGill Univ., Montreal, QC H3A 2T8, Canada; \url{parente@physics.mcgill.ca}}
\altaffiltext{2}{National Radio Astronomy Observatory, Charlottesville, VA 22903, USA}
\altaffiltext{3}{Dept.~of Physics, Concordia Univ., Montreal, H4B 1R6, Canada}
\altaffiltext{4}{National Research Council of Canada, Herzberg Astronomy and Astrophysics, Dominion Radio Astrophysical Observatory, P.O. Box 248, Penticton, BC V2A 6J9, Canada}
\altaffiltext{5}{Dept. of Astronomy, Cornell Univ., Ithaca, NY 14853, USA}
\altaffiltext{6}{Center for Advanced Computing, Cornell Univ., Ithaca, NY 14853, USA}
\altaffiltext{7}{Dept.~of Physics and Astronomy, West Virginia Univ., Morgantown, WV 26506, USA} 
\altaffiltext{8}{Center for Gravitational Waves and Cosmology, West Virginia University, Chestnut Ridge Research Building, Morgantown, WV 26505}
\altaffiltext{9}{National Astronomical Observatories, Chinese Academy of Science, 20A Datun Road, Chaoyang District, Beijing 100012, China}
\altaffiltext{10}{Max-Planck-Institut f\"{u}r Radioastronomie, Auf dem H\"{u}gel 69, D-53121 Bonn, Germany}
\altaffiltext{11}{Leibniz Universit ̈at Hannover, D-30167 Hannover, Germany}
\altaffiltext{12}{Max-Planck-Institut f̈ur Gravitationsphysik, D-30167 Hannover, Germany}
\altaffiltext{13}{Columbia Astrophysics Laboratory, Columbia Univ., New York, NY 10027, USA}
\altaffiltext{14}{Center for Advanced Radio Astronomy, Univ. of Texas Rio Grande Valley, Brownsville, TX 78520, USA }
\altaffiltext{15}{SKA South Africa, Pinelands, 7405, South Africa}
\altaffiltext{16}{Dept. of Physics and Astronomy, Franklin and Marshall College, Lancaster, PA 17604-3003, USA}
\altaffiltext{17}{George Mason University, resident at the Naval Research Laboratory, Washington, DC 20375, USA}
\altaffiltext{18}{Faculty of Science, Univ. of East Anglia, Norwich Research Park, Norwich NR4 7TJ, United Kingdom} 
\altaffiltext{19}{ASTRON, Netherlands Institute for Radio Astronomy, Postbus 2, 7990 AA, Dwingeloo, The Netherlands}
\altaffiltext{20}{Anton Pannekoek Institute for Astronomy, Univ. of Amsterdam, Science Park 904, 1098 XH Amsterdam, The Netherlands}
\altaffiltext{21}{Center for Gravitational Wave Astronomy, Univ. Texas Rio Grande Valley - Brownsville, TX 78520, USA}
\altaffiltext{22}{Physics Dept., Univ. of Wisconsin - Milwaukee, 3135 N. Maryland Ave., Milwaukee, WI 53211, USA}
\altaffiltext{23}{Jodrell Bank Centre for Astrophys., School of Phys. and Astro., Univ. of Manchester, Manch., M13 9PL, UK}
\altaffiltext{24}{Dept. of Physics and Astronomy, Univ. of British Columbia, Vancouver, BC V6T 1Z1, Canada}
\altaffiltext{25}{NRAO, PO Box 0, Socorro, NM 87801, USA}

\graphicspath{{figures/}}
\begin{abstract}
The Pulsar Arecibo L-Band Feed Array (PALFA) survey, the most sensitive blind search for radio pulsars yet conducted, is ongoing at the Arecibo Observatory in Puerto Rico. The vast majority of the 180 pulsars discovered by PALFA have spin periods shorter than 2 seconds. Pulsar surveys may miss long-period radio pulsars due to the summing of a finite number of harmonic components in conventional Fourier analyses (typically $\sim$16), or due to the strong effect of red noise at low modulation frequencies. We address this reduction in sensitivity by using a time-domain search technique: the Fast-Folding Algorithm (FFA). We designed a program that implements a FFA-based search in the PALFA processing pipeline, and tested the efficiency of the algorithm by performing tests under both ideal, white noise conditions, as well as with real PALFA observational data. In the two scenarios, we show that the time-domain algorithm has the ability to outperform the FFT-based periodicity search implemented in the survey. We perform simulations to compare the previously reported PALFA sensitivity with that obtained using our new FFA implementation. These simulations show that for a pulsar having a pulse duty cycle of roughly 3\%, the performance of our FFA pipeline exceeds that of our FFT pipeline for pulses with DM $\lesssim$ 40 pc cm$^{-3}$ and for periods as short as $\sim$500 ms, and that the survey sensitivity is improved by at least a factor of two for periods $\gapp$ 6 sec. Early results from the implementation of the algorithm in PALFA are also presented in this paper.
\end{abstract}

\section{Introduction}
One characteristic of the population of known radio pulsars is that 93$\%$ of them have spin periods (P) shorter than 2 seconds\footnote{Based on the ATNF Pulsar Database, version 1.56} (\url{https://www.atnf.csiro.au/people/pulsar/psrcat/)}. The notable lack of long-period pulsars could be an intrinsic property of the population. For instance, the observed population of slowly rotating pulsars (defined here as having P $>$ 2 sec) have radio beam widths smaller than typical pulsars. Indeed, the median pulse duty cycle, $\delta$, defined as the ratio of the full width at half maximum (FWHM) of the pulse to the pulsar period, for this class of pulsars is 1.6$\%$, while it is 3.1$\%$ for pulsars with spin periods shorter than 2 sec. The beaming of the radiation would therefore play a role in the detectability of slow pulsars. The lower spin-down luminosity of long-period pulsars is another factor that could explains why these pulsars are particularly difficult to detect. \\ 

In addition to effects that are intrinsic to the pulsars themselves, the lack of long-period pulsars in the known population may also be due to selection bias in pulsar surveys. One of the reasons why surveys are likely to miss slowly rotating pulsars is that pulsar search radio data are often badly affected by red noise, or excess noise at low modulation frequencies. This non-Gaussian noise is the result of the combined effects of various factors such as receiver gain fluctuations and radio frequency interference (RFI). The broad features introduced in the time series by red noise increase the number of false positives in the low modulation frequency regime (defined in this paper as $f$ $<$ 0.5 Hz), where red noise is strongest, causing a considerable reduction in the sensitivity of pulsar surveys at this end of the spectrum. For the Pulsar Arecibo L-Band Feed Array (PALFA) survey, the fact that the integration time of the observations is only 268 sec is another limiting factor of the detectability of the survey to long-period pulsars. \\

While Fourier-based search techniques have been commonly used in blind searches for pulsars, their performance are highly compromised by red noise. By recovering synthetic pulsar signals injected in real observational data with $\texttt{PRESTO}$'s \citep{r01} Fast-Fourier Transform (FFT) search program ($\texttt{accelsearch}$), \citet{lbh+15} demonstrated that there are major discrepancies between the true sensitivity of the PALFA survey and the sensitivity predicted by the radiometer equation \citep{dtws85}. For a hypothetical pulsar having a spin period of 10 sec and a DM of 10 pc cm$^{-3}$, the minimum mean flux density the FFT can detect is 20 times larger than the value predicted by the radiometer equation. The degradation in the true sensitivity is noticeable at pulsar spin periods as short as few hundreds of milliseconds. \\

One way to partially address this reduction in sensitivity is by the use of a Fast-Folding Algorithm (FFA, \citealt{s69}), a time-domain search technique especially well-suited for finding long-period signals. The FFA folds a dedispersed time series at multiple trial periods and avoids redundant summation of bins by storing in memory the resulting sum of each folding step, and later reusing these stored quantities when needed. The main advantage of the FFA over a frequency-domain search is that by producing a phase-coherent result, it retains all harmonic structure, as opposed to FFT-based searches where only a limited number of harmonics\footnote{A maximum of 32 summed harmonics is used in the case of the PALFA \texttt{PRESTO}-based pipeline; see \citet{lbh+15}}. are incoherently summed (i.e., without using the phase information in the harmonics). Having a search technique that is efficient at finding narrow-pulsed signals in the long-period regime is therefore desirable. \\ 

Recovering the loss in sensitivity reported in \citet{lbh+15} is important as it has the potential for scientific advancements in pulsar astronomy. Our understanding of the Galactic pulsar population is heavily biased by various selection effects. These include the propagation effects in the interstellar medium, the non-uniform radio sky background, the distances and the proper motions of pulsars, as well as the sizes of the emission beams. In addition to the observational effects mentioned above, red noise also likely affects the observed period distribution of the pulsar population. Finding more slowly rotating pulsars would help us constrain the radio emission mechanism: one of the longest-period radio pulsars known \citep{ymj99}, PSR J2144$-$3933 (P = 8.5 sec), challenges existing models as this object is located beyond the theoretical death line \citep{cr93,zhm00,ha01} in the P$-\dot{\textrm{P}}$ diagram. The very recent discovery of a 23.5-sec pulsar in the LOFAR Tied-Array All-Sky Survey\footnote{\url{http://www.astron.nl/lotaas/}}, PSR J0249+58 (Tan et al., in prep.), further motivates the search for long-period pulsars. Furthermore, optimizing our detection capabilities at low modulation frequencies increases the chances of discovering the first neutron star - black hole binary system. Since the black hole will presumably have resulted from the supernova explosion of the initially more massive star in the binary, a pulsar companion will not have been recycled and so would generally have similar periods to the non-recycled pulsar population \citep{ppr05,lba05,e07}. Such a discovery could provide valuable insights into stellar evolution and serve as a testbed for theories of gravity. Increased sensitivity to low modulation frequencies also makes pulsar surveys more likely to find radio-loud magnetars: the four known radio-loud magnetars have rotation periods between 2 and 6 sec (see e.g., \citealt{kb17}). \\

The use of the FFA has been fairly limited over the past decades. \citet{ls69} implemented the algorithm when working at the Arecibo Observatory, resulting in the discovery of PSR B2016+28 ($P=0.56$ sec, \citealt{csc68}). The Parkes Multibeam Pulsar Survey used the FFA to search for periodic signals in the data collected by the survey, which led to the discovery of the 7.7-sec pulsar J1001$-$5939 \citep{fsk+04,lfl+06}. It was also used in a search for radio pulsations in observations of the 6.85 sec X-ray pulsar XTE J0103$-$728, but resulted in no significant detections \citep{cldtk09}. \citet{kml+09} used the FFA to perform a search for periodicity on radio observations of six X-ray dim isolated neutron stars (XDINSs) and then compared the sensitivity of the time-domain algorithm to that of a typical Fourier-based technique. This work demonstrated the ability of the FFA to exceed the performance of the FFT in the white noise regime, especially when searching for pulsars having high harmonic content. \citet{cbc+17} recently obtained results similar to those presented by \citet{kml+09}, where an in-depth study of the behavior of the time-domain algorithm was conducted in both a Gaussian noise regime and in real observational data collected by the High Time Resolution Universe (HTRU) pulsar survey. This analysis showed an enhancement in the detectability of long-period pulsars when using the FFA in the two regimes. The use of the algorithm also extends exoplanet hunting, which is similar to pulsar searches, only dips are observed in the time series rather than pulses. It was used to search for transits by Earth-size planets around G and K-type dwarfs stars in Kepler data \citep{pmh13} and it led to the discovery of a number of exoplanet candidates. \\

Deploying a FFA-based search in a large-scale pulsar survey is computationally expensive, and this is the main reason why the use of this alternative technique has been limited in the past. Nevertheless, the increasing power of modern supercomputers allows us to use the FFA in a large-scale pulsar survey. \\ 

In this paper, we present the results from the implementation of a FFA-based search, \texttt{ffaGo}\footnote{Available at \url{https://github.com/emilieparent/ffaGo}}, in the PALFA survey. We compare the efficiency of \texttt{ffaGo} to that of a FFT pulsar searching program in both the ideal, white noise regime and in real PALFA survey data. The expected sensitivity of the FFA in the large-scale PALFA survey is evaluated by reproducing an analysis similar to that presented in \citet{lbh+15}, where various pulsar signals are injected in a selection of PALFA observations files free of astrophysical signals and then recovered using \texttt{ffaGo} to determine the minimum mean flux density our FFA-based pipeline can detect in the PALFA survey. \\

The organization of this paper is as follows: Section 2 offers a brief mathematical description of the FFA. Details regarding the implementation of the algorithm and the testing of significance metrics used to evaluate FFA-generated profiles in the PALFA survey are discussed in Section 3. In Section 4, we compare the performance of the FFA to that of the FFT using both simulated and real data collected at Arecibo containing long-period pulsars. We then report on the sensitivity analysis conducted with the FFA in Section 5, where we recover synthetic pulsar signals injected in real PALFA data. Section 6 presents the results from the implementation of the time-domain algorithm in PALFA, along with new discoveries made by the survey. Finally, we summarize the main results of this paper in Section 7.
\section{The Fast-Folding Algorithm}
The FFA was originally developed by \citet{s69} for searching for periodic signals in the presence of noise in the time domain, in contrast to the Fast Fourier Transform search technique which operates in the frequency domain. By avoiding redundant summations, the FFA is much faster than standard folding at all possible trial periods: it performs summations through $ N \log_{2}(N/p - 1) $ steps rather than $ N(N/p - 1)$, where $N$ and $p$ are the number of samples in the time series and the trial folding period in units of samples, respectively. Large computational power is still required when applying the FFA over a very wide range of trial periods and this is why the use of the FFA in large-scale pulsar searches has been limited in the past.\\

The FFA folds each dedispersed time series with sampling interval $\Delta t$ at multiple periods ($p$, in units of sample time), and our implementation of the algorithm then looks for statistically significant features in the generated profiles. The algorithm performs partial summations, while avoiding redundancy, into a series of $\log_{2}p$ stages and then combines those sums in different ways so that the data are folded with a trial period between $p$ and $p+1$. A time series containing $N$ time samples folded in a FFA execution at the folding period $p$ (corresponding to a period in time units of P $= p \times \Delta t$) will result in $ M = N/p$ different pulse profiles with slightly different periods ranging from $p_i$ to $p_i + 1$: 
\begin{equation}
  p_{i} = p_0 + \Big(\frac{i}{M-1}\Big)   
\end{equation}
where $p_0$ is the effective folding period and 0 $\leqslant$ i $\leqslant M -1$. \\

\begin{figure}[hb!]
  \includegraphics[width=\columnwidth]{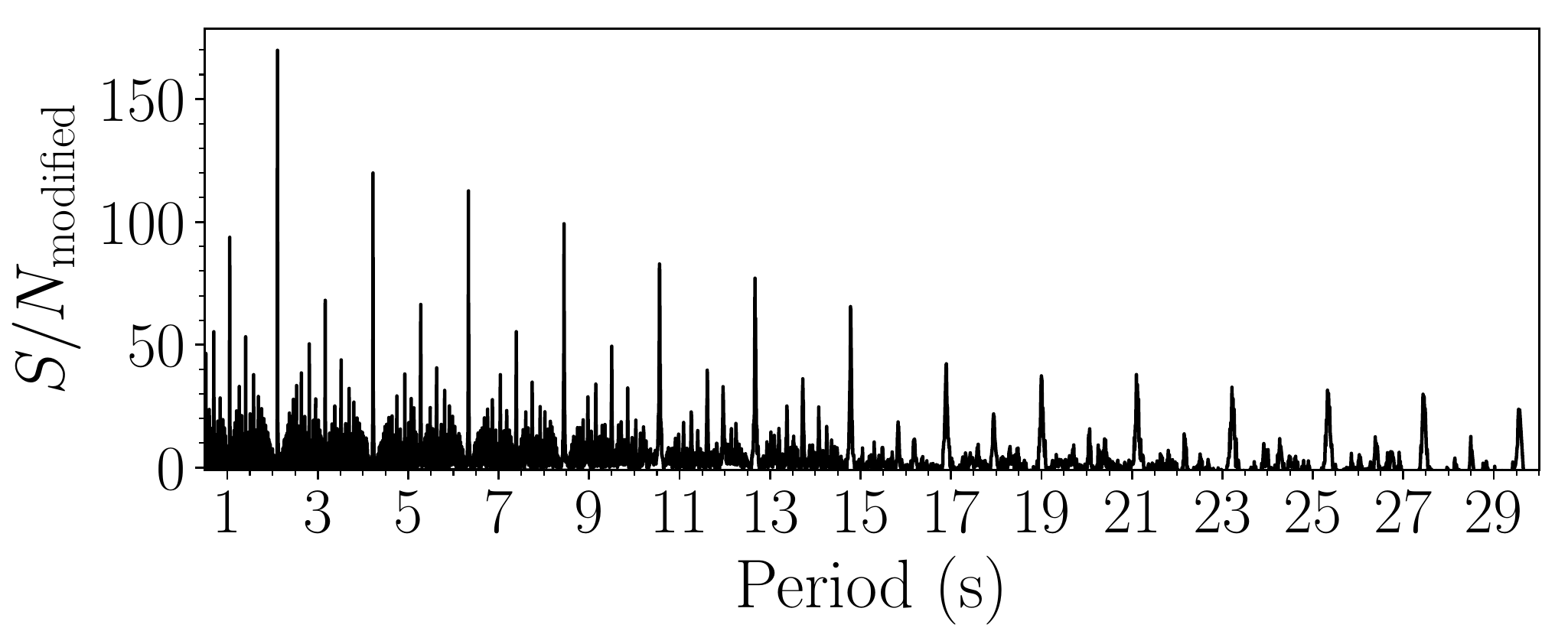}
  \caption{Periodogram of PSR J2004+3137 generated by \texttt{ffaGo}. One can clearly identify the fundamental period of the pulsar (P = 2.11 sec), as well as many harmonics.}
  \label{fig:periodo}
\end{figure}

\begin{figure*}[ht!]
    \includegraphics[scale=0.56]{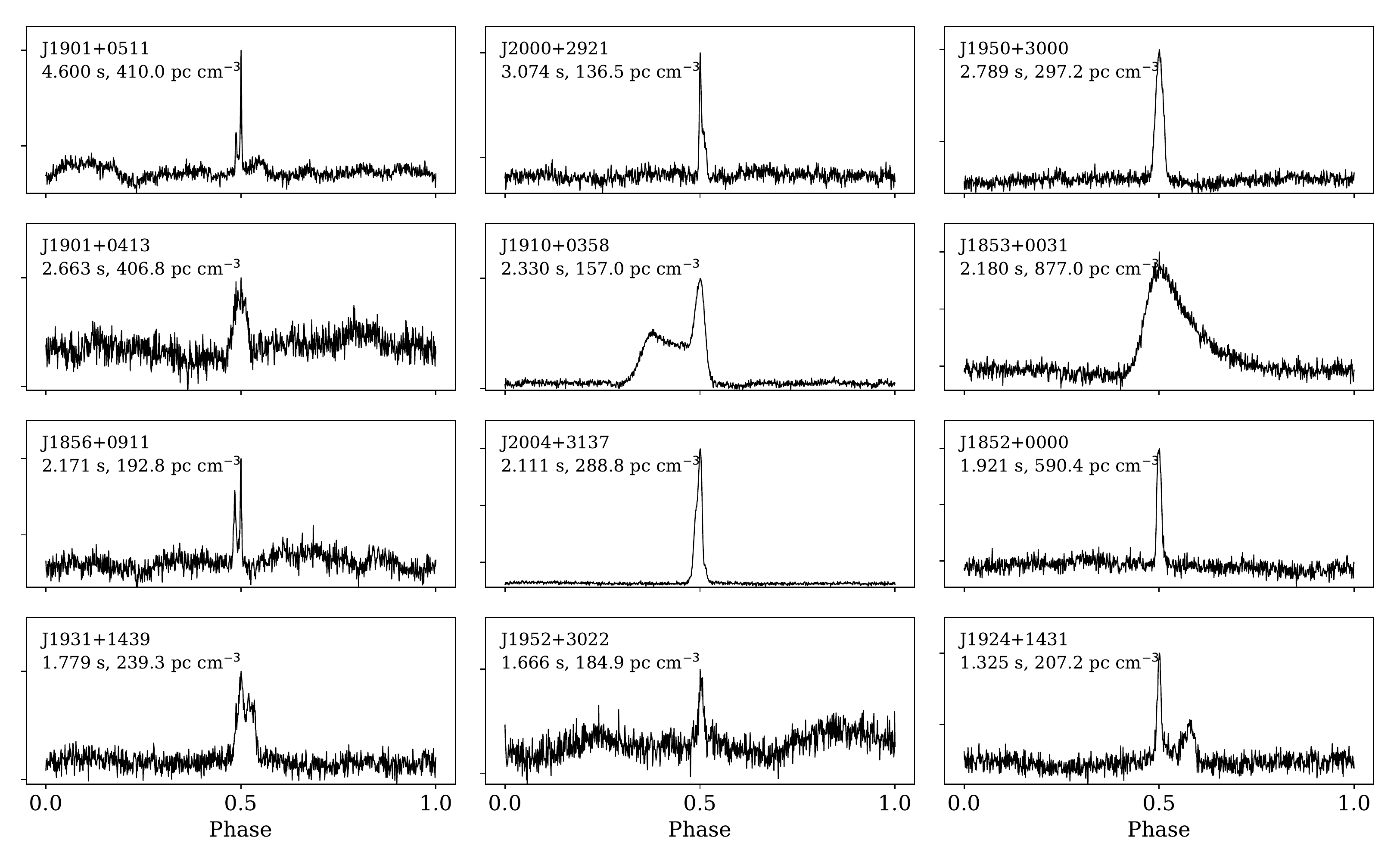}
    \caption{Pulse profiles of 12 long-period pulsars discovered with the \texttt{PRESTO}-based PALFA pipeline in  observations with the Mock spectrometer at 1.4 GHz. The profiles were folded using \texttt{PRESTO}'s \texttt{prepfold} program.} The name, period and DM of the pulsars are specified above each profile. One can see the broad features in the baseline introduced by red noise and interference in the data, especially prominent for PSRs J1901+0413, J1856+0911, and J1952+3022.
    \label{fig:profs}
\end{figure*}

While the folding procedure is a core component of a FFA-based search, the statistical evaluation of the resulting profiles is another crucial component of the search. This is discussed in Section 3.3. Figure~\ref{fig:periodo} shows an example periodogram one obtains from applying the FFA on a 268-sec PALFA observation of the bright, long-period pulsar J2004+3137, when looking for periodicity between 500 ms to 30 sec. The pulse profile of this source is shown in Figure~\ref{fig:profs}. The peak in S/N is at the pulsar's fundamental period, 2.11 sec, and the secondary peaks are the harmonics and sub-harmonics of the spin period. 
The FFA requires $\log_2(N/p)$ to be an integer, or equivalently, $M$ to be a power of 2. If this condition is not satisfied, our implementation of the algorithm will pad the time series by its median value. A more complete description of the FFA algorithm can be found in \citet{s69}, \citet{ls69} and \citet{lk04}. \\

The main advantages of the FFA over the FFT are that the FFA offers greater frequency resolution (especially important in the low-frequency end of the spectrum) and, most importantly, that it coherently sums all harmonics of a signal (i.e., it folds the data in phase). Indeed, the incoherent harmonic summing that is used in Fourier-domain searches inevitably misses power in higher harmonics, since one must choose a finite number of harmonics to be summed when using this technique. Hence, the FFA is more sensitive to narrow pulses. 
\section{A FFA-Based Pipeline in the PALFA Survey}
PALFA has two independent search pipelines performing a full-resolution analysis: the primary \texttt{PRESTO}-based pipeline \citep{lbh+15} and the Einstein@Home-based pipeline \citep{akc+13}. A reduced-resolution analysis is also performed on-site at the Arecibo Observatory: this “Quicklook” pipeline \citep{sjsk+13} allows rapid discovery and confirmation of bright pulsars. The work presented here will focus only on the \texttt{PRESTO}-based pipeline, which has been modified to additionally perform the FFA-based search for long-period pulsars. \\

This pipeline runs on the Guillimin Supercomputer, part of McGill University's High Performance Computing centre operated by Compute Canada and Calcul Qu$\acute{e}$bec. PALFA data are transferred from the Arecibo Observatory to the Cornell University Centre for Advanced Computing (CAC), from where they are downloaded to Guillimin. The results from the data processing pipeline are uploaded upon completion to the PALFA database, also located at the CAC, for future human inspection. \\ 

In the \texttt{PRESTO}-based pipeline, the 4-bit data files (\texttt{PSRFITS} format) are first subject to RFI mitigation routines. The data are then dedispersed at a wide range of trial dispersion measures (DMs). A Fourier-based periodicity search is subsequently performed on the dedispersed time series using \texttt{PRESTO}'s \texttt{accelsearch} software. The pipeline also has a single pulse search component \citep{Patel:Thesis:2016} that searches for single, dispersed pulses up to DM values of 10 000 pc cm$^{-3}$ (Patel et al., in prep). The PALFA Consortium then uses the online collaborative tool on the CyberSKA platform\footnote{\url{https://ca.cyberska.org/}} \citep{kab+11} to classify generated pulsar and transient candidates. For more details about PALFA's data processing, see \citet{lbh+15}.

\subsection{Implementation of the FFA in the PALFA pipeline}
We have designed a Python program, \texttt{ffaGo}\footnote{Available at \url{https://github.com/emilieparent/ffaGo}}, that implements the FFA-based periodicity search into the PALFA analysis software. \texttt{ffaGo} reads any 32-bit float time series produced by \texttt{PRESTO}, and it includes a de-reddening procedure aimed at reducing the effect of red noise on the input time series. This de-reddening is done by applying a dynamic median filter, where the size of the filtering window is, by default, set to twice the largest trial period searched. To shorten FFA executions, downsampling of the data is performed initially such that the sampling interval is approximately 2 ms. The data are then normalized by dividing by the maximum value before calculating the standard deviation, $\sigma$, of the time series for future profile evaluations (see Section 3.1.2). Subsequent dynamical rebinning routines are carried out to search for multiple pulse widths. \\ 

Parts of our FFA code are taken from an open-source FFA package\footnote{Available at \url{https://github.com/petigura/FFA}}, written as a Python and C program developed for transit searches in Kepler data \citep{pmh13}. More specifically, the parts of our code that wrap the time series, pad it, and perform the folding and the summations were taken from \citet{pmh13}.\\

Signal-to-noise ratio (S/N) calculations, candidate selection and sifting are also incorporated in this program. Periodograms similar to Figure~\ref{fig:periodo} can also be generated by \texttt{ffaGo}. We note that the primary focus while designing the CPU-based $\texttt{ffaGo}$ was not to minimize the computation time. Large scale, real-time analyses should consider parallelized versions of FFA-based searches.
\subsection{Search parameters}
The pulsar parameter space that we consider in the implementation of \texttt{ffaGo} in the pipeline consists of the following:
\begin{enumerate}[label=\Alph*.]
\item \textit{The Period} \\ We search for periods ranging from a minimum of 500 ms to a maximum of 30 sec. Even though the FFA is designed to be fast, it is still computationally expensive to apply to higher modulation frequencies, since they produce a large number of profiles that need to be statistically evaluated. This in turn results in an important increase in the computational burden: searching down to 100 ms nearly doubles the time required to process one time series with \texttt{ffaGo}. This is one of the reasons why the blind search is restricted to periods longer than 500 ms. Moreover, \citet{lbh+15} demonstrated that 500 ms is approximately the period at which one notices a decrease in the sensitivity of PALFA at low DMs. It is not worth looking for periods larger than 30 sec with \texttt{ffaGo} since the integration time of PALFA observations are of 268-sec and 180-sec for the inner (32$^\circ \lesssim l \lesssim 77^\circ$) and outer (168$^\circ \lesssim l \lesssim 214^\circ$) Galaxy regions, respectively: it is unlikely that folding fewer than $\sim$ 10 pulses will result in significant detections, especially in the presence of red noise. We rely on the single pulse search conducted in the pipeline to identify pulses from very slow (P $>$ 30 sec) pulsars (Patel et al., in prep).\\

\item \textit{The Pulse Width} \\ To explore the pulse width parameter while optimizing the S/N and minimizing the computation time, we perform rebinning by a factor of \rm{X} at multiple stages during the search such that the sampling interval ranges from $\sim$2 ms up to a few seconds, depending on the trial period and the pulse duty cycle {$\delta$} we are searching for at each step of the process. The PALFA time series, which initially have a sampling interval of 65 $\mu$s are first decimated so that each bin has a width of approximately 2 ms. Afterwards, we divide the 500 ms - 30 sec full range of trial periods into six sub-ranges, processed separately, such that the fixed sampling interval is no smaller than 1/1000 of the shortest trial period in the sub-range and no larger than 1/100 of that period. In other words, the minimum $\delta$ we search is kept between 0.1$\%$ and 1$\%$, when assuming a pulse fully enclosed within one bin. We impose this lower limit on the searched range of pulse widths to reduce the execution time. Additional rebinning is applied to the time series before entering FFA executions in each sub-range to ensure that the ratio of the sampling interval to the shortest trial period is greater than 1/1000. Further downsampling of the time series is performed within each period sub-range in order to efficiently search $\delta$ values ranging from approximately 0.2-0.5$\%$ up to 10-13$\%$. The downsampling factors we use are 2$^k$ and 3$^k$, where 1 $\leqslant k \leqslant$ 3. To ensure optimal sensitivity, this last downsampling stage is performed at different phases (i.e., adjacent bins are summed in different ways). \\
\item \textit{The DM} \\ Since the values of DM of the pulsars to be discovered are unknown, a large number of DM trials must be used in the search. We search with the FFA from DM = 0 pc cm$^{-3}$ to DM = 3265 pc cm$^{-3}$ in steps of 5 pc cm$^{-3}$, resulting in 653 dedispersed time series to be processed through \texttt{ffaGo}. Using finer DM steps is unnecessary as we are searching in the long-period phase-space, where the pulse widths are typically from a few to hundreds of milliseconds. The only scenario where our sensitivity could be affected by this coarse DM spacing is one where a pulsar had a value of DM that sits exactly between two trial DMs, which corresponds to a dispersive smearing of 2.6 ms, and if that particular pulsar had a short spin period and a narrow pulse width (for example, shorter than 500 ms and a pulse duty cycle smaller than 0.5\%). We are searching up to DMs higher than the maximum Galactic value predicted by NE2001 \citep{cl02}, which is about 2000 pc cm$^{-3}$ in the region surveyed by PALFA, to account for any possible dense, local regions that could be not included in the model. The DM step size was chosen such that the amount of processing is minimized while avoiding sensitivity loss from channel smearing due to dispersion. We are not searching above DM = 3265 pc cm$^{-3}$ since the probability of finding normal pulsars with mean flux densities of a few mJy outside our Galaxy observation is quite low considering the relatively short integration time of PALFA observations (see Section 3). 
\end{enumerate}
\subsection{Profile evaluation}
The significance metric that we use to evaluate profiles generated by the algorithm assumes that the profile has one single-peaked pulse, that this pulse is constant in phase and that it is captured within a single bin (i.e., the detection is optimal when the bin size is equal to the width of the profile). The mathematical description of the metric (Metric A) is as follows:
\begin{equation}
  \textrm{S/N} = \frac{ I_{\rm max} - I_{\rm med} }{(\sigma\sqrt{\rm{X}}) \, \sqrt{M - z}},
\end{equation}
where ${I}_{\rm max}$ and $I_{\rm med}$ are the maximum and the median intensities of the folded profile, and $\sigma$ is the standard deviation of the time series, calculated after the initial downsampling, detrending and normalization of the time series. Subsequent rebinning is accounted for by multiplying the standard deviation by the square-root of the downsampling factor, \rm{X}. Finally, $z$ is the fraction of a profile that requires padding such that the necessity of the number of profiles $M$ being a power of two is respected. \\

We also explore other metrics for evaluating profiles, such as one in which the median and the standard deviation would be calculated only over the off-pulse portion of the profile, so that the on-pulse component is not included when statistically characterizing the baseline noise in each profile. \citet{kml+09} and \citet{cbc+17} used such a metric to evaluate profiles generated by a FFA program\footnote{The respective programs can be downloaded from \url{https://github.com/vkond/ffasearch} and \url{https://github.com/adcameron/ffancy}}. Specifically, we tested Metric B, where we exclude a 20$\%$ window centered on the peak of the profile when calculating the median, $I_{\rm med,off}$, and the standard deviation, $\sigma_\mathrm{off}$, of the profile. As opposed to Metric A, in which the denominator of the expression for the S/N is constant for a given FFA execution, the standard deviation $\sigma_{\rm off}$ in Metric B is calculated directly on the off-pulse portion of individual profiles produced within a FFA execution. While using this algorithm to evaluate FFA-generated profiles, we explore the pulse width phase-space by applying the downsampling procedure described previously, rather than a boxcar matched-filtering approach \citep{cm03}, as was done in \citet{kml+09} and in \citet{cbc+17}. The S/N of the peak in each FFA-generated profile is then calculated as follows:
\begin{equation}
  \textrm{S/N} = \frac{{I}_{\rm max} -I_{\rm med,off}}{\sigma_{\rm off}}.
\end{equation}\\

To compare the efficiency of Metric A and B, we performed a search using both metrics on a dataset of simulated pulsar signals constructed with \texttt{SIGPROC}'s\footnote{\url{https://github.com/SixByNine/sigproc}} \texttt{fake} program, which injects periodic top-hat pulses in Gaussian noise. The synthetic pulsars have spin periods, P, ranging from 2 to 20 sec (in increments of 2 sec), with pulse duty cycles $\delta$ of 0.5$\%$, 1$\%$, and from 2$\%$ to 20$\%$ with step size of 2$\%$, resulting in 120 different trial combinations of period/pulse width. Each of these trials was constructed and tested five times to ensure that no statistical anomalies were introduced in our dataset when using the \texttt{fake} program. In total, 600 data files were searched with both metrics. The amplitude of individual pulses, $S$, was chosen such that the total pulse energy, $E = \textrm{P} S \delta $, was kept fixed for each trial. Broader pulses therefore have lower peak fluxes compared to narrow pulses. The sampling interval of the fake observations was set to 65 $\mu$s with a 268-sec integration time at a central observing frequency of 1375 MHz and a bandwidth of 322 MHz to match real PALFA data when observing inner Galaxy regions. The DM value at which all signals were injected was arbitrarily chosen to be 150 pc cm$^{-3}$. \\

\begin{figure}[ht!]
    \includegraphics[width=\columnwidth]{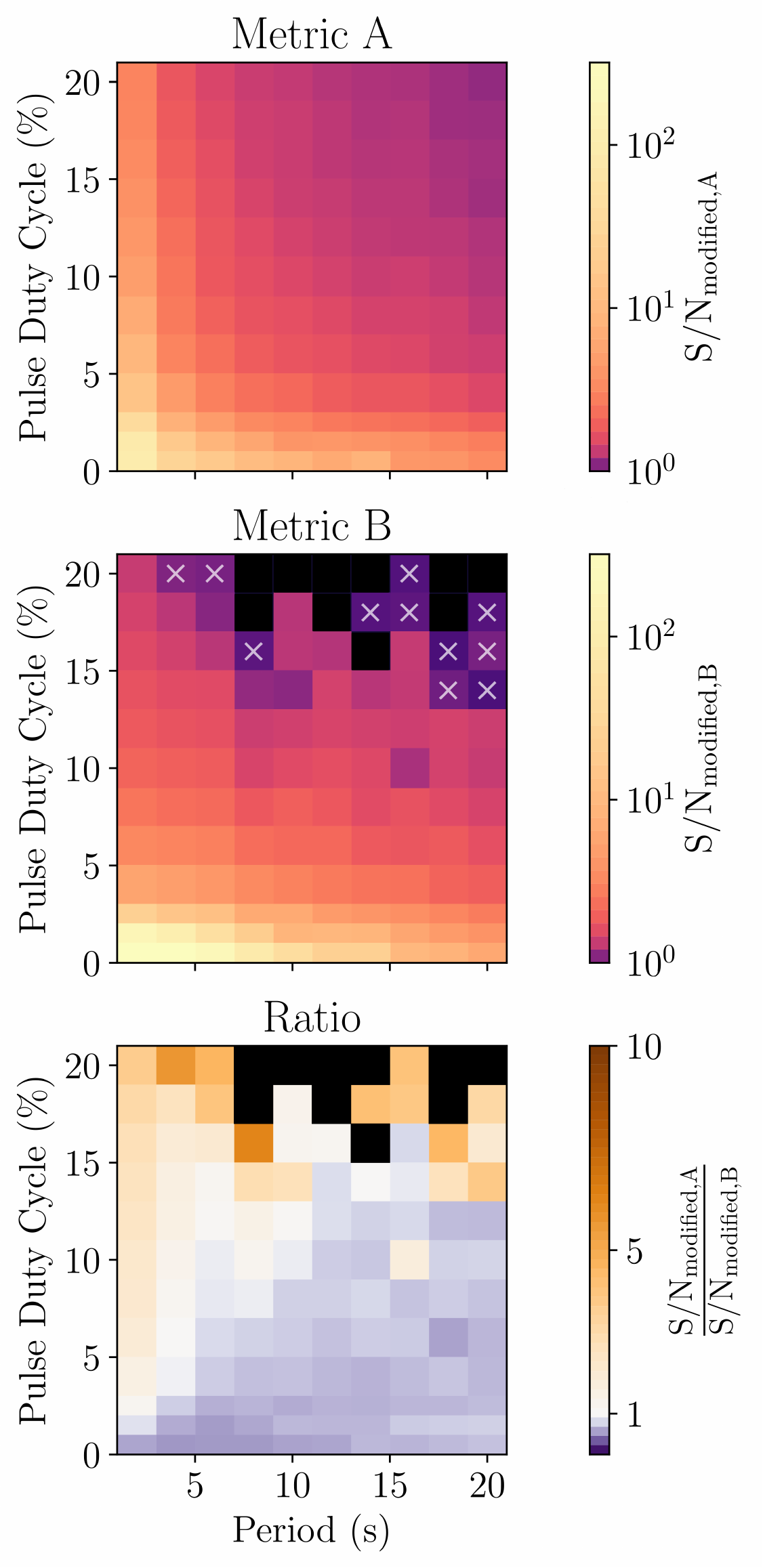}
    \caption{Response patterns of the two FFA significance metrics, Metric A (\textit{top panel}) and Metric B (\textit{middle panel}), investigated in the white noise simulation described in Section 3.3. The ratio of the two values of $\rm S/N_{\rm modified}$ is shown in the \textit{bottom panel}. The values reported are the average $\rm S/N_{\rm modified}$ from the five simulations. Black pixels represent trials that were not detected in all five datasets, while pixels with white crosses represent those having an average $\rm S/N_{\rm modified}$ below 6 (i.e., trials that were classified as non-detections).}
    \label{fig:SNR_B}
\end{figure}

The simulated observation files were dedispersed at the appropriate DM prior to searching periodicities between 500 ms and 30 sec with both metrics. Once the search was completed, the lists of candidates were inspected by eye to identify the highest $\rm S/N_{\rm modified}$ values (Section 3.4 describes how $\rm S/N_{\rm modified}$ differs from S/N) at which the artificial pulsars were detected. The results of this simulation are shown in Figure~\ref{fig:SNR_B}. \\

The response pattern from Metric A shows that it provides the best detections for narrow-pulsed, short-period signals. The optimal detection occurs at the shortest trial period of 2 sec and $\delta=0.5\%$. The $\rm S/N_{\rm modified}$ values then gradually fall off. This is expected because, for longer periods/wider profiles, the amplitude of the pulse is reduced since we require the total pulse energy to remain constant. \\

For Metric B, the response pattern suggests that the determining factor when it comes to the metric responsiveness is the pulse width: this metric responds strongly to narrow profiles and its sensitivity decreases only slightly with increasing period. Moreover, this metric reaches higher $\rm S/N_{\rm modified}$ values for the trials with narrow pulse widths compared to Metric A. Metric B remains significantly responsive up to $\delta = $ 8-12$\%$, above which it practically vanishes. This behavior is also shown in the bottom panel of Figure~\ref{fig:SNR_B}, where we see that, for all periods, Metric B is outperformed by Metric A at large values of pulse duty cycle $\delta$. Figure~\ref{fig:SNR_B} also suggests that Metric A is better at detecting signals with short periods (P$\lesssim$ 4 sec) and $\delta$ larger than $\sim 2-5\%$. However, we also see that Metric B yields larger $\rm S/N_{\rm modified}$ values than Metric A for narrow-pulsed signals having long periods.  \\

One clear distinction between the two metrics is that Metric A detected all artificial pulsars, while ten trials having broad profiles were missed by Metric B in all five simulations (black pixels in Figure~\ref{fig:SNR_B}). Furthermore, 11 trials were detected by Metric B with an average $\rm S/N_{\rm modified}$ below the threshold for candidate folding set in the pipeline, meaning that we consider those trials as being not successfully detected by Metric B. Therefore, 21 out of 120 fake pulsars were not detected by Metric B.  \\

\citet{cbc+17} also investigated a significance metric similar to Metric B when evaluating FFA-generated pulse profiles, and concluded that even if such a metric possesses the ability to outperform the FFT in the long-period regime, it suffers from sensitivity deterioration when it comes to broad pulses. This characteristic can however help in reducing the number of false positives generated by red noise in the data. The analysis presented here is consistent with the results presented in \citet{cbc+17}, and demonstrates that it is likely that the survey would miss pulsars having broad profiles if this metric were used in the FFA search. For an interpretation of the difference in the performance of the two metrics, see the Appendix. \\

We also designed an alternative, Metric C, which, similarly to Metric B, excludes a 20\% window centered on the peak to calculate the median intensity of the profile, $I_{\rm med,off}$. The standard deviation of Metric C is similar to that used in Metric A, only we include an extra factor of $\sqrt{0.8}$ in the profile's standard deviation to account for the on-pulse exclusion (see Eq. 5 in the Appendix). The same set of synthetic pulsars injected in white noise described above was searched with Metric C. Results from this analysis suggest that Metric C has a response pattern very similar to Metric A, and that there is no significant difference between the two metrics. Unlike Metric B, Metric C suffers negligible loss in sensitivity for large $\delta$ values. Therefore, we conclude that Metric A and Metric C are equivalent. More details on profile evaluation with Metric C, including the response pattern obtained from the white noise simulation, can be found in the Appendix.\\

Due to the non-detection of wider pulses by Metric B, we opted for implementing Metric A to evaluate FFA-generated profiles in the PALFA processing pipeline, which successfully detected all trials and showed a response pattern that suggests overall broader sensitivity. We note that Metric C would also have been a reasonable option. When downloading \texttt{ffaGo}, the user can select any of the three metrics described in this work. \\

\subsection{Candidate selection}
For each dedispersed time series processed through the pipeline, all FFA-generated profiles are statistically evaluated (see Section 3.3) to identify periodic signals. 
A set of S/N values (i.e., a periodogram) is produced each time we downsample the initial time series at a specific phase (i.e., at each possible way of summing adjacent bins) by a factor of $2^k$ or $3^k$, as described in Section 3.2 B. These sets have different statistical distributions, because the number of profiles generated for a specific period will vary as the number of samples in the rebinned time series changes. To avoid being biased in the candidate selection process, we make the S/N sets uniform by subtracting the mode of that S/N value's distribution, and then dividing by its median absolute deviation (MAD): 
\begin{equation}
  \rm S/N_{\rm modified,i} = \frac{\rm S/N_i - \rm mode_i}{\rm MAD_i}, 
\end{equation}
where i represents a specific set of S/Ns (i.e. the periodogram obtained for a specific rebinned time series). All candidates are therefore characterized by a modified S/N value, $\rm S/N_{\rm modified}$, which estimates the significance of the S/N calculated by the selected metric. The mode and the MAD were chosen for their robustness when evaluating statistics of largely  skewed distributions,  as is the case when pulsar signals are present in the data. \\

All candidate periods detected with a $\rm S/N_{\rm modified}$ $\geqslant$ 5 are recorded to a list along with the $\rm S/N_{\rm modified}$, the sampling interval and the value of DM at which the candidate was detected. This is done for all 653 dedispersed time series and the full FFA search uses approximately 10$\%$ of the PALFA pipeline total processing time, which corresponds to a few hours. The set of candidate lists are subsequently sifted using a modified version of \texttt{PRESTO}'s \texttt{sifting} routine, also included in the open-source \texttt{ffaGo} package. This sifting removes weaker, harmonically related periods and RFI-like signals and groups candidates according to their DM. More details regarding the general candidates sifting procedure can be found in \citet{lbh+15}. \\

Once the time series have been searched and the FFA candidates have been sifted, only candidates having $\rm S/N_{\rm modified}$ $\geqslant$ 6 are selected for folding. This limit is also applied to the candidates produced by \texttt{accelsearch} in the PALFA pipeline to reduce the number of false positives that have to be inspected. The raw data are folded with \texttt{PRESTO}'s \texttt{prepfold} routine at each candidate period. Similarly to FFT-generated candidates, we do not allow \texttt{prepfold} to search in period and DM space if the candidate has a period greater than 500 ms to avoid converging to nearby RFI. The resulting plots, along with ratings calculations \citep{lbh+15} and one rating from a candidate-ranking artificial intelligence (AI) system (\citealt{zbm+14}), are then uploaded to a PALFA's online Candidate Viewer application for final human inspection and classification. FFA-generated candidates generally represent approximately 10$\%$ to 25$\%$ of the total number of folded periodicity candidates, which varies between 150 to 250 total candidates per beam. \\

\section{Comparing the FFA to the FFT}
\subsection{Comparison Using Simulated Data}
\begin{figure}[ht!]
    \includegraphics[width=\columnwidth]{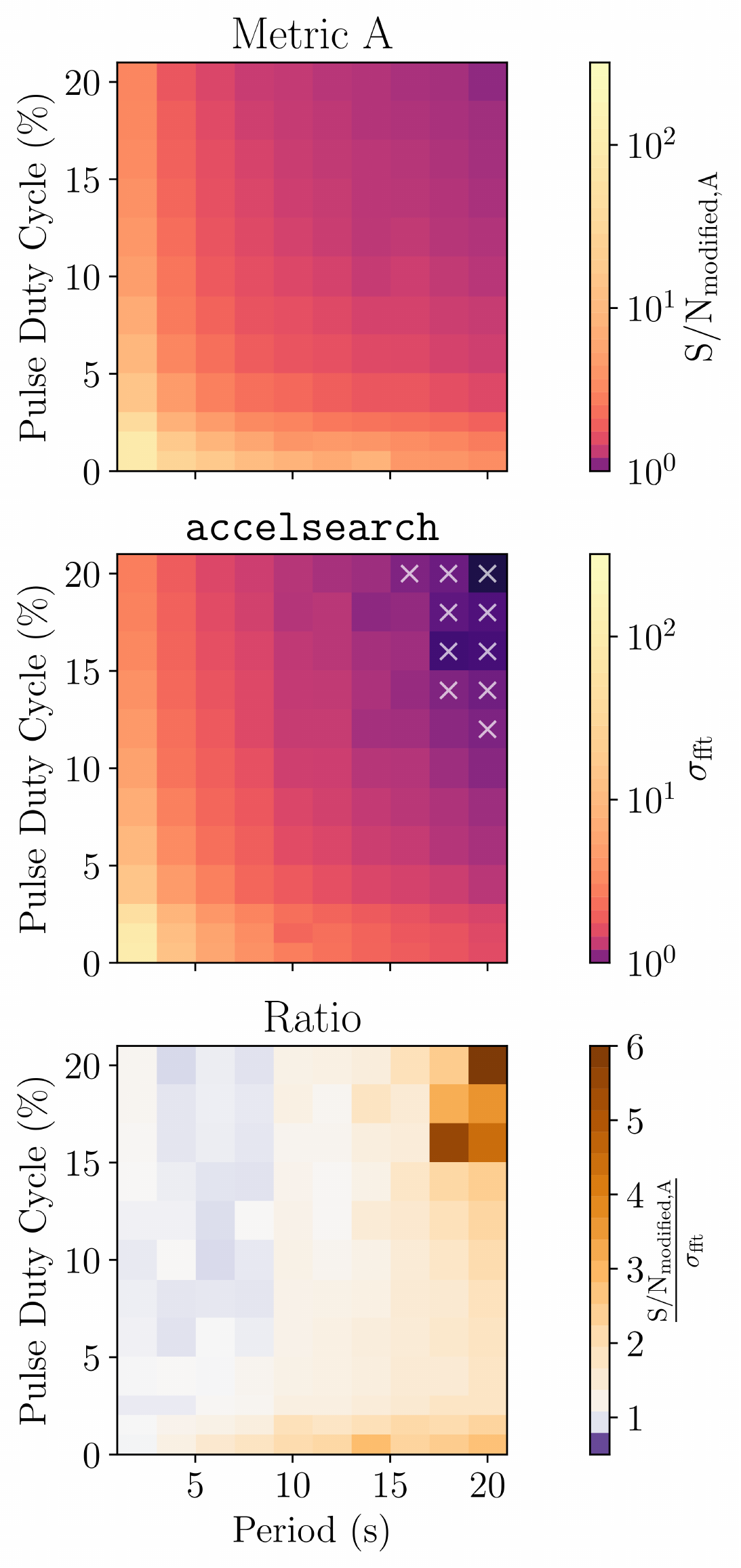}
    \caption{Response patterns of the FFA when using Metric A (\textit{top panel}) and \texttt{accelsearch} (\textit{middle panel}), investigated in the white noise simulation described in Section 3.3. Pixels with white crosses represent those having an average $\rm \sigma_{\rm fft}$ below 6. The \textit{bottom panel} represents the ratio of the $\rm S/N_{\rm modified}$ over the $\sigma_{\rm fft}$ for each trial. Although the numerical value of the ratios do not reflect directly the sensitivity gain achieved by the FFA, it allows us to visualize where the improvement is maximal. The values reported are the average $\rm S/N_{\rm modified}$ from the five simulations. Note that the scale for the top and the middle panels are logarithmic, while the bottom panel is displayed on a linear scale.}
    \label{fig:FFAvsFFT}
\end{figure}

To compare the performance of the \texttt{ffaGo} program to that of a typical Fourier-based search, \texttt{PRESTO}'s \texttt{accelsearch} program was applied to the five datasets of 120 artificial pulsar signals that were used in the analysis presented in Section 3.3. The Fourier-based search summed up to 32 harmonics incoherently, and the significance of the FFT candidates were characterized by a $\sigma_{\rm fft}$ value, the quantity used in the PALFA survey to evaluate the strength of a FFT candidate. The value of $\sigma_{\rm fft}$ is determined by calculating the equivalent Gaussian significance of the candidate based on the probability that the same amount of incoherently summed power is noise. In the PALFA pipeline, candidates with $\sigma_{\rm fft}$ values greater than 2 are recorded to a list of candidates that are later sifted, but only candidates with $\sigma_{\rm fft}$ above 6 are folded and uploaded to the online Candidate Viewer for human inspection. Therefore, we consider here only signals having $\sigma_{\rm fft}$ $\geqslant$ 6 as successfully detected by the program. The $\rm S/N_{\rm modified}$ from the FFA search (Metric A) and the $\sigma_{\rm fft}$ from \texttt{accelsearch} at which the simulated pulsars were detected were recorded for the two periodicity searches, and the strength of the detections are illustrated in Figure~\ref{fig:FFAvsFFT}. It is important to note that the type of statistics used to characterize the detections made by the algorithms are fundamentally different. Therefore, numerical scores from the two searches should not be directly compared. \\

Both FFA and FFT searches show similar response patterns with similar regions of maximum sensitivity: even under ideal white noise conditions, the detected S/N$_{\rm modified}$ values decrease with increasing period and increasing pulse width (i.e., decreasing peak amplitude). This is expected since we require the per-pulse energy to be constant and there are fewer pulses in the 268-sec time series when injecting longer periods. The response from the frequency-domain algorithm however falls off more sharply with period as compared to the time-domain search.  \\

A major difference that arises between the two techniques is that, while the FFA successfully recovered all trials, \texttt{accelsearch} detected 10 trials (pixels with white crosses in Figure~\ref{fig:FFAvsFFT}) showing broad profiles with an average $\sigma_{\rm fft}$ value below 6 (some of these trials were totally missed by \texttt{accelsearch}). These are not considered as successful detections since such candidates would have been excluded from the final list of potential candidates generated by the processing pipeline. While we expect the FFT to be particularly sensitive to signals having low harmonic content, the lowest modulation frequencies are effectively searched via their highest harmonics and, in the PALFA processing pipeline, \texttt{accelsearch} searches down to a minimum of 1 Hz. The program is therefore intrinsically less sensitive to very long-period pulsars having low harmonic content. This restriction on the lowest frequencies searched is set in order to reduce the number of false positive candidates produced by red noise in the data. This explains why the algorithm is outperformed by the FFA in the broad pulse regime and why some trials were missed by the frequency-domain search. \\

The bottom panel in Figure~\ref{fig:FFAvsFFT} shows the ratio of the $\rm S/N_{\rm modified}$ to the $\sigma_{\rm fft}$ values. The resulting pattern can be used to illustrate the phase-space where the use of the FFA is the most advantageous. Although the two numerical scores cannot be compared directly due to the fundamental difference in their nature, the displayed pattern suggests that there are two particular regions where the FFA is more responsive. First, we see that the coherent summing of all harmonics makes the time-domain algorithm more efficient at finding the pulsar signals having the smallest pulse widths, and this advantage grows with increasing period. The second region is where trials have the broadest pulses and the lowest spin frequencies. We emphasize once more the arbitrary nature of the values of ratio shown in Figure~\ref{fig:FFAvsFFT}, especially considering the fact that the two quantities compared do not scale equivalently to increasingly bright signals.  \\

In summary, this analysis demonstrated the ability of the FFA to outperform the frequency-domain search in the long-period regime in the presence of white noise. Similar studies were carried out by \citet{kml+09} and by \citet{cbc+17} and also demonstrated that, even if every trial were detected by the FFT, the performance of a FFA exceeds that of a FFT. We also showed that a FFT can fail to detect broad signals with P $>$ 18 sec even in ideal conditions for a 268-sec integration time. This shows that even in the absence of red noise, the coherent summing of all harmonics is necessary to detect some long-period pulsars. \\

A similar simulation is presented in Section 6, where artificial pulsars have been injected in real observational data rather than in white noise to quantify the efficiency of the FFA when searching for pulsars in a large-scale survey under real RFI conditions. \\

\subsection{Comparison Using Real Pulsar Data}

\begin{table*}
  \caption{Results from the analysis performed on 12 long-period pulsars discovered by the PALFA survey.}
  \setlength{\tabcolsep}{0.55cm}
  \begin{tabular}{c c c c c c c }
  \hline
  \multicolumn{1}{c}{PSR name} & \multicolumn{1}{c}{Period} & \multicolumn{1}{c}{Pulse duty cycle} & \multicolumn{3}{c}{FFA $\rm S/N_{\rm modified}$} & \multicolumn{1}{c}{FFT}\\ 
    \cline{4-6}
    {} & (s) & ($\%$ of phase)& Metric A & Metric B & Metric C& $\sigma_{\rm fft}$ \\
    \hline
    \hline
    J1901+0511 &  4.600 	& 0.4  & 33.3 & 35.8 & 24.8	 & 12.0$^\diamond$ \\  
    J2000+2921 &	3.074	& 0.8  & 55.0 & 63.4 & 47.1	 & 18.5$^\diamond$\\
    J1950+3000 &	2.789	& 2.2  & 78.9 & 92.2 & 72.7 	 & 12.1	\\
    J1901+0413 &	2.663 	& 3.1  & 8.5$^\diamond$  & 9.7$^\diamond$  & 7.2$^\diamond$ 	 & 9.4	 \\
    J1910+0358 &	2.330	& 3.3  & 47.3 & 12.8 & 14.5$^\diamond$ 	 & 31.3	\\
    J1853+0031 &	2.180	& 9.7  & 84.1 & 19.4 & 76.3   & 68.1	\\
    J1856+0911 &	2.171 	& 0.5  & 21.7 & 24.3 & 17.7	 & 8.3$^\diamond$	 \\
    J2004+3137 &	2.111	& 1.6  & 169.5 & 163.4 & 157.9	 & 104.4	 \\
    J1852+0000 &	1.921	& 1.3  & 59.5 & 55.0 & 54.4	 & 34.7	\\
    J1931+1439 &	1.779	& 1.6  & 65.2 & 72.8 & 52.27	 & 22.1$^\diamond$\\ 
    J1952+3022 &	1.666	& 1.0  & 15.1 & ...  & 13.2	 & 7.8$^\diamond$	 \\
    J1926+0431 &	1.325	& 1.1  & 53.1 & 51.2 & 47.9 	 & 27.0 \\
    \hline 
  \end{tabular}
  \begin{tablenotes}
   \small
   \item $^\diamond$ Detection via an harmonic of the fundamental frequency of the pulsar. 
  \end{tablenotes}
  \label{tab:long-p_pulsars}
\end{table*}

To evaluate the response of \texttt{ffaGo} to pulsars in the presence of RFI and red noise, and compare it to a FFT-based search, we applied the program to a dataset of 12 PALFA observations collected at the Arecibo Observatory containing a variety of long-period pulsars discovered by the survey \citep{slm14,lbh+15,lsb+17}. We then compared the significance of the detections from the FFA search against that obtained by \texttt{accelsearch}. We also processed the dataset through the FFA using Metrics B and C to evaluate their responses in presence of red noise. \\

The selected observations contained pulsar signals covering a period range from 1.32 sec to 4.6 sec and having $\delta$ values ranging from less than 1$\%$ up to $\sim$10$\%$. The pulse duty cycle, $\delta$, values reported in Table~\ref{tab:long-p_pulsars} were measured by calculating the fraction of bins with intensity larger than half the maximum value in the integrated pulse profiles. While most of the sources display single-peaked profiles, some pulsars from our dataset have two-component profiles (see profiles in Figure~\ref{fig:profs}). For example, PSRs J1901+0511 and J1856+0911 both exhibit two narrow closely spaced pulse components, while PSR J1924+1431 has a broad and a narrow component that are separated in phase. We were also interested in quantifying the detectability of pulsars having broad profiles, such as PSRs J1852+003 and J1910+035, in the red noise regime. When considering the width of the entire pulse (i.e., the portion of the profile around the peak that is above the baseline intensity), the on-pulse fraction for these two sources are 30.5$\%$ and 21.7$\%$, respectively (but they have $\delta$ of 9.7$\%$ and 3.3$\%$, respectively, when they are calculated via their pulse FWHM).\\ 

Prior to dedispersion of the \texttt{PSRFITS} observation files at the appropriate DMs of the pulsars, the data were cleaned of interference by applying \texttt{PRESTO}'s \texttt{rfifind} routine, which identifies narrow-band RFI and produces a mask for bad time and frequency intervals. To optimize detections, we produced time series dedispersed at multiple DM values around the true DM of the pulsars. \\

We processed each masked and dedispersed time series through \texttt{ffaGo} (searching for periodicities ranging from 500 ms to 30 sec) as well as through \texttt{PRESTO}'s \texttt{accelsearch} to search in the Fourier domain while incoherently summing up to 32 harmonics. The candidate periodicities from both searches were then separately sifted and the lists of final candidates were then inspected by eye to identify the strongest candidates harmonically related to the pulsar. \\

Metrics A and C, as well as the FFT search, were successful at detecting the full selection of pulsars (see results in Table~\ref{tab:long-p_pulsars}). Except for PSR J1901+0413 for which the detection was marginal ($\rm S/N_{\rm modified}$ $\lesssim$ 10), the $\rm S/N_{\rm modified}$ of the FFA detections with Metric A were all well above the threshold $\rm S/N_{\rm modified,i} \geqslant$ 6 that we consider for candidate folding, meaning that the regular processing pipeline would have folded the pulsars for final human classification. All pulsars were detected at their fundamental frequency when using Metric A in the FFA search, while there were five instances where the FFT search detected pulsars via their harmonics. Moreover, four of the detections made with \texttt{accelsearch} were marginal detections ($\sigma_{\rm fft}< 10$). The important conclusion we draw from this analysis is that, even if the numerical scores of the detections made by both FFA and FFT searches cannot be compared directly, the FFA successfully recovered the true period of a variety of pulsars with different pulse profiles (some having multiple components), at $\rm S/N_{\rm modified}$ values significantly larger than the detectability threshold set in the PALFA pipeline. The FFT search detected a number of these sources at harmonics of their spin frequencies and in some cases only marginally. \\

When using Metric B, there were two cases where the source was missed. Interestingly, it was neither the longest-period nor largest $\delta$ pulsars that were missed. From their pulse profiles we can see that the non-detected pulsars are the ones that have baselines showing broad features introduced by red noise. This further motivates the choice of Metric A over Metric B for FFA-generated profile evaluation in the pipeline implementation of the algorithm. In general, the significance of the detections with Metric C are marginally lower than Metric A's, and two pulsars were detected at a harmonic of the fundamental frequency. This suggests that Metric A is slightly more efficient in presence of red noise. 

\section{Sensitivity of the PALFA survey}
To assess the true sensitivity of the PALFA survey, artificial pulsar signals were constructed and injected in real survey data using \texttt{PRESTO}'s \texttt{injectpsr} (described in \citealt{lbh+15}). This program generates smeared, scattered and scaled pulse profiles that are added to real data at regular time intervals corresponding to the chosen spin period. To scale the profiles properly, observations of the radio galaxy 3C 138 (for which measurements of the flux density are available in the literature) were carried out in December 2013. During these observations, a calibrating noise diode was turned on so that the flux density of the diode could be compared to that of the galaxy. Per-channel scaling factors between flux density and the observation data units were then calculated (see \citealt{lbh+15} for more details on the calibration procedure). These scaling factors are used to obtain the targeted phase-averaged flux density ($S_{\rm mean}$) of the artificial pulsar signals. \\

Signals constructed with \texttt{injectpsr} have single von Mises \citep{von18} component pulse profiles with FWHM specified by the user. Dispersive smearing and scattering are then applied to the profile, where the amount of broadening caused by scattering (in ms) is determined by the specified value of DM and the observing frequency, $\nu$ (in GHz), according to the following \citep{bccnl04}:
\begin{multline}
    \log{\tau_{\rm scat}} = -6.46 + 0.154\log{\textrm{DM}}\\+ 1.07(\log{\textrm{DM}})^2 - 3.86\log{\nu}.
\end{multline}
The modified data are then recorded into a \texttt{SIGPROC} ``filterbank'' file format. \texttt{injectpsr} can then be used to create a dataset of synthetic pulsars for which we can adjust $S_{\rm mean}$ and then characterize the sensitivity of a survey at each point in (period, DM, pulse FWHM) phase-space. More details regarding the construction of synthetic pulsars with \texttt{injectpsr} can be found in \citet{lbh+15}. 

\subsection{Sensitivity of the PALFA Survey using a Fourier-based Search Technique}
A realistic sensitivity analysis of the \texttt{PRESTO}-based PALFA pipeline was conducted in \citet{lbh+15} to evaluate the true performance of \texttt{accelsearch} at finding diverse types of pulsars in PALFA data.\\

An important result from \citet{lbh+15} is that there is a clear mismatch between the PALFA survey sensitivity curves measured and the ideal case predicted by the radiometer equation \citep{dtws85} when searching in the long-period regime. At low DMs, a reduction in the survey sensitivity is noticeable for spin periods as short as $\sim$100 ms. At a spin period of $\sim$11 sec, a pulse FWHM of 2.6\%, the measured $S_{\rm min}$ values are 10 and 20 times larger than the predicted value for dispersion measures of 10 pc cm$^{-3}$ and 600 pc cm$^{-3}$, respectively. \\

\citet{lbh+15} also injected synthetic pulsar signals in Gaussian noise. The measured curves in this scenario also rise at longer periods, although the degradation in sensitivity is not as pronounced as in the real data injections case. At a period of $\sim$11 sec and DMs of 10 pc cm$^{-3}$ and 600 pc cm$^{-3}$, while the minimum detectable average flux densities measured from real data injections are approximately 10 to 20 times larger than the values predicted by the radiometer equation, the minimum detectable average flux densities measured from white noise data injections are still 3 to 5 times larger than the predictions. This indicates that RFI and red noise alone cannot explain the discrepancy between the measurements and the predictions, and that the periodicity searching component of the pipeline is subject to potential improvement. 

\subsection{Sensitivity of the PALFA Survey using the FFA Search}
\begin{figure*}[ht!]
  \begin{center}
  \includegraphics[scale=0.63]{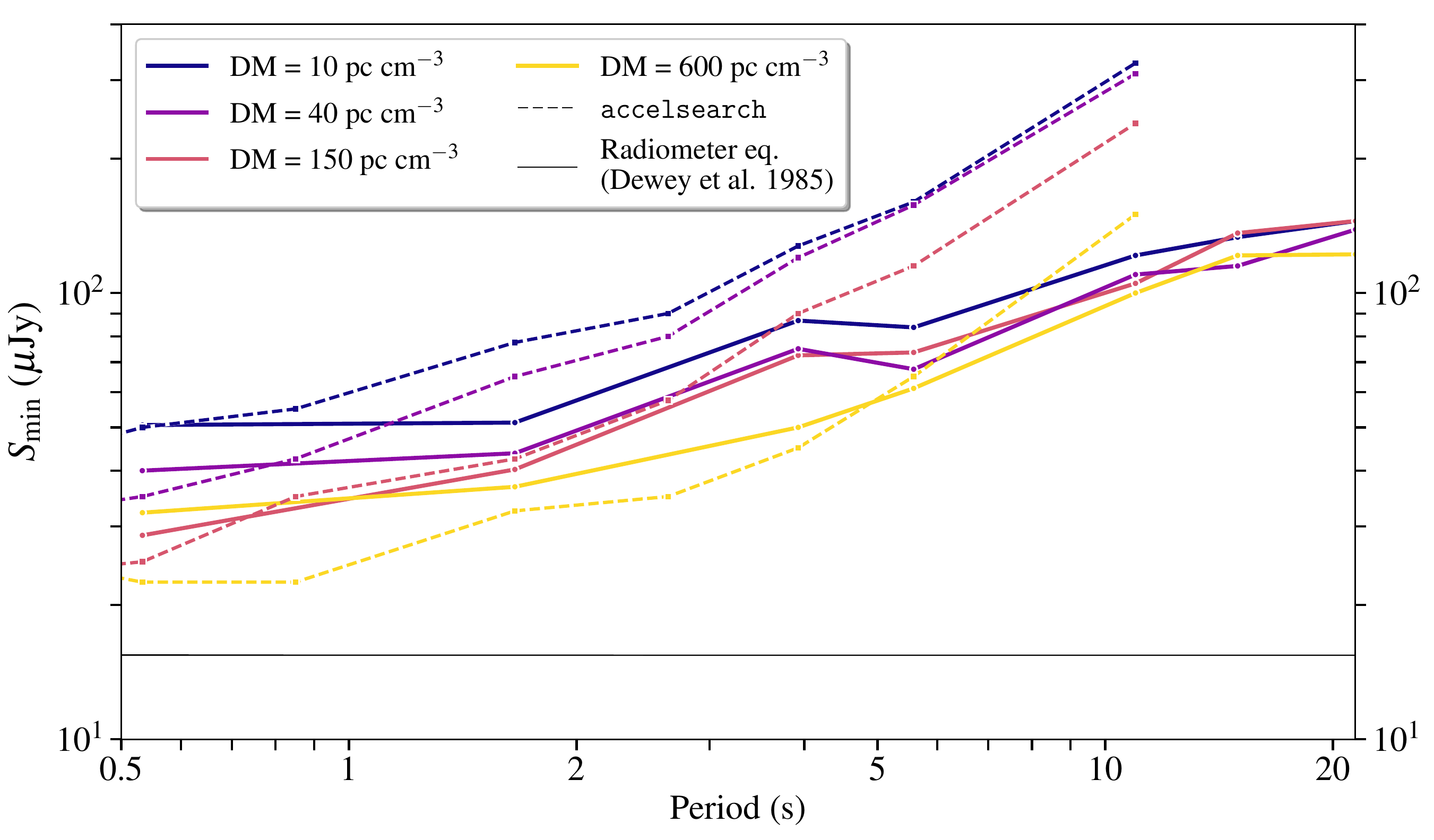}
  \caption{Minimum detectable mean flux density for the PALFA survey as measured when searching synthetic pulsar signals injected in real data with \texttt{ffaGo}. The signals have a fixed pulse FWHM of 2.6$\%$. $S_{\rm min}$ measured by the FFA are illustrated with solid lines, while the dashed lines represent values of $S_{\rm min}$ obtained from the frequency-domain search reported in \citet{lbh+15}. Note the greater parameter space covered at long periods in the FFA analysis.}
  \label{fig:sensit}
  \end{center}
\end{figure*} 
\begin{figure}
  \includegraphics[width=\columnwidth]{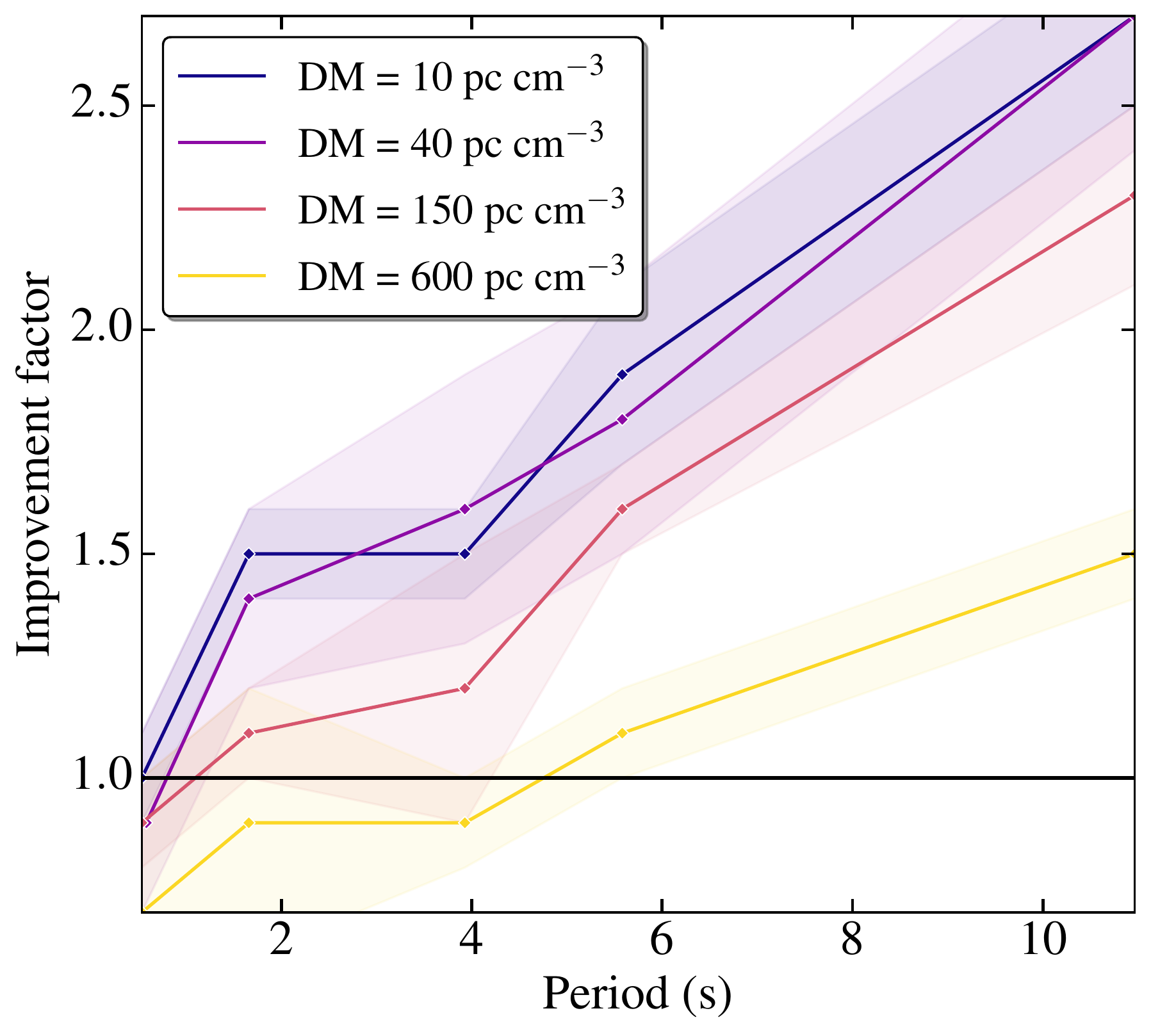}
  \caption{Improvement factor in the survey sensitivity as a function of period for the 4 DM trial values used, where the pulse FWHM = 2.6$\%$ is kept fixed. The factors were determined by dividing the median value of $S_{\rm min}$ from \citet{lbh+15} for a given trial by the $S_{\rm min}$ obtained from the \texttt{ffaGo}-based search for that trial. The shaded regions represent the uncertainty on the improvement factors, which were derived from the difference between the minimum detectable flux densities and the largest value of mean flux densities for which the trials were missed by the searches.}
  \label{fig:improv}
\end{figure} 
We reproduced the analysis described above to assess the sensitivity of the survey in the long-period phase-space when using our implementation of the FFA (Metric A) to search for pulsar signals in PALFA data. Synthetic pulsars with periods longer than 500 ms were injected into the same real survey data that were used in \citet{lbh+15}, and four of the DM trial values used in the previous analysis were selected. To avoid confusion with RFI, the trial periods were chosen to be non-integer values. We extended the period parameter space up to periods of $\sim$15 and $\sim$21 sec to evaluate the responsiveness of the FFA in the long-period regime. Synthetic pulsars with FWHM pulses of 0.5$\%$, 1.5$\%$, 2.6$\%$, 5.9$\%$ and 11.9$\%$ were injected into the PALFA datasets, and only the signals with a pulse FWHM of 2.6$\%$ were injected in all 12 data files. The complete list of pulsar parameters used in the work presented here can be found in Table~\ref{tab:b}. \\

The minimum detectable mean flux density required for the FFA to detect the injected signals having a FWHM of 2.6$\%$ is shown in Figure~\ref{fig:sensit}. Similarly to the FFT-based search, the FFA sensitivity curves do not flatten out at longer periods, as opposed to what the radiometer equation predicts \citep{dtws85}. However, the degradation in sensitivity is not as pronounced as the FFT curves reported in \citet{lbh+15}, implying a gain in sensitivity: the FFA outperforms the FFT search for periods as short as 550 ms at DM = 10 pc cm$^{-3}$. The performance of the FFA search does not seem to vary as strongly with DM at the longest periods compared to the FFT, which has a stronger response to pulsars having large values of DM. Figure~\ref{fig:improv} illustrates the factors of improvement in the sensitivity resulting from using the FFA versus the FFT-based search. As expected, the gain in sensitivity is greater in the longer periods/low DM phase space.\\

\begin{figure*}[ht!]
  \includegraphics[scale=0.44]{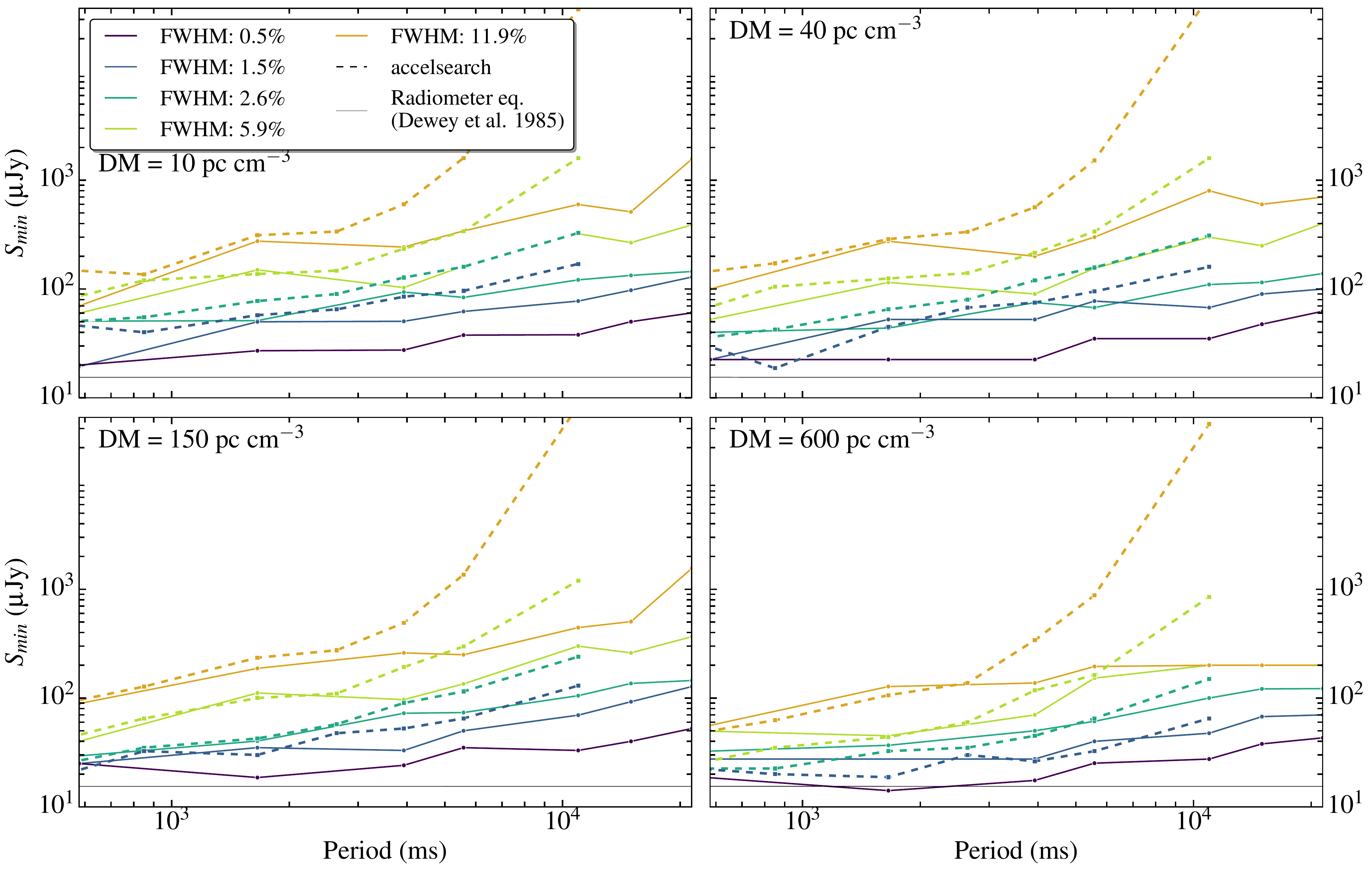}
  \caption{Minimum detectable mean flux density for the PALFA survey, as measured when searching synthetic pulsar signals injected in real data with \texttt{ffaGo}, for various pulse widths. Each panel corresponds to a different value of DM. Results from the FFA analysis presented in this work are illustrated with solid lines, while dashed lines represent the results reported in \citet{lbh+15}. Note that the injections for the FFA analysis included narrow pulse profiles having full width at half maximum of 0.5$\%$ (purple lines) that were not included in \citet{lbh+15}}.\\ \\
  \label{fig:sensit_pw}
\end{figure*} 
\begin{figure*}[ht!]
  \includegraphics[scale=0.40]{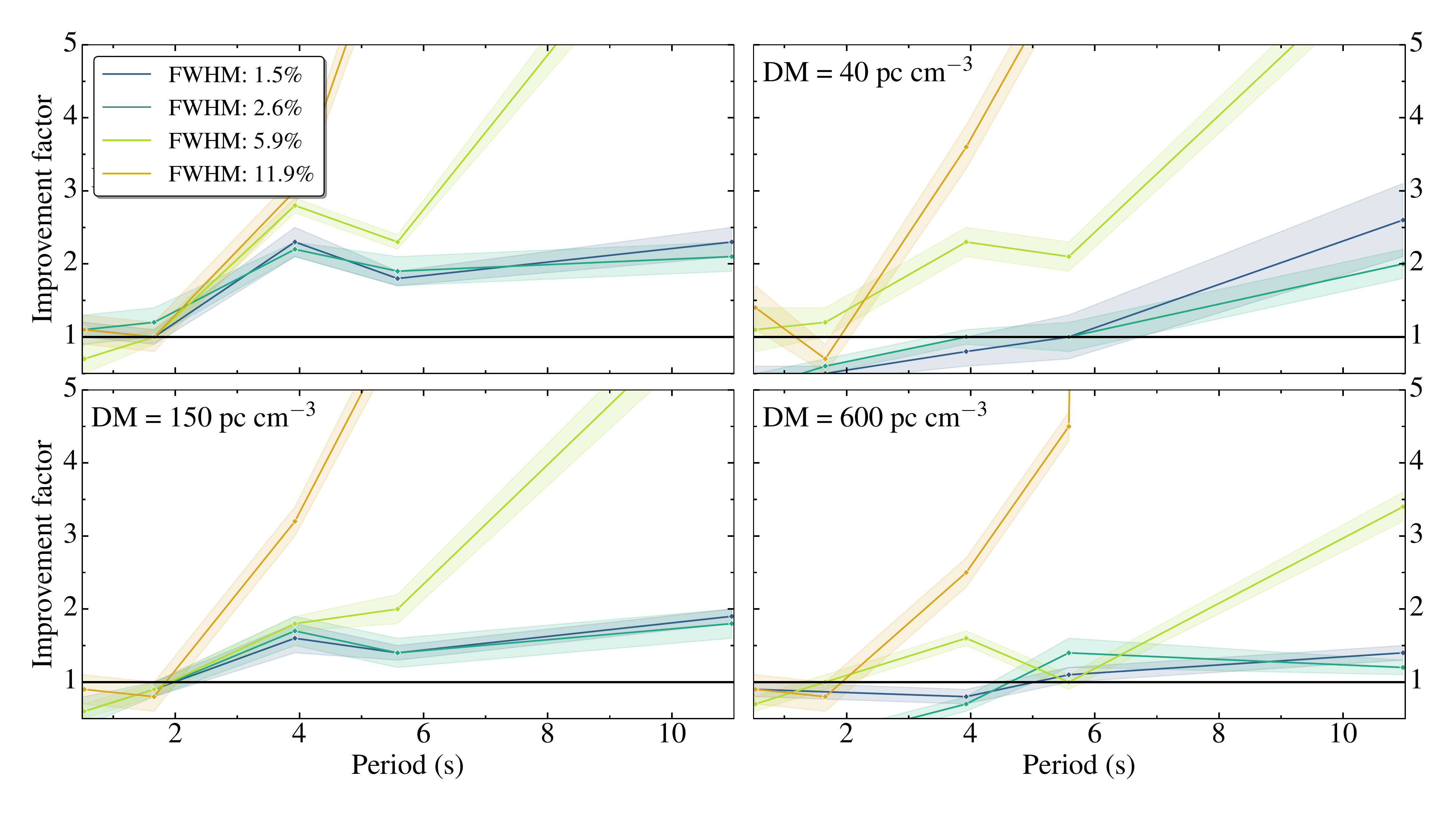}
  \caption{Improvement factor in the survey sensitivity as a function of period for the four DM trial values used and for various pulse FWHMs. The factors were determined by dividing the value of $S_{\rm min}$ from \citet{lbh+15} for a given trial by the $S_{\rm min}$ obtained from the \texttt{ffaGo}-based search for that trial. The shaded regions represent the uncertainty on the improvement factors, which were derived from the difference between the minimum detectable flux densities and the maximum value of mean flux densities for which the trials were missed by the searches.}
  \label{fig:improv_pw}
\end{figure*} 
The second part of the analysis consisted of injecting signals with different pulse FWHM in one of the observational data files. One can see from the results shown in Figures~\ref{fig:sensit_pw} and ~\ref{fig:improv_pw} that, for periods greater than $\sim$ 8 sec, the FFA is more efficient than the FFT for all values of DM and all pulse FWHMs. The gain in sensitivity is much greater when searching for pulsars having broad profiles. The advantage of a time-domain search over a frequency-domain search is that the coherent summing of all harmonics makes the search especially responsive to signals having narrow pulse profile, while a Fourier transform should be more sensitive to profiles with low harmonic content. Recovering low-modulation frequencies is however more difficult if red noise obscures the fundamental frequency, as well as low harmonic frequencies, which may contain a considerable fraction of the total power in the case of pulsars with wide profiles. In addition to the deterioration due to red noise, \texttt{accelsearch} searches very low modulation frequencies via the highest harmonics of a particular frequency, as mentioned in Section 4.1. This explains partially the important gain in sensitivity observed for broad pulses shown in Figures~\ref{fig:sensit_pw} and ~\ref{fig:improv_pw}, and confirms the results from the white noise simulation presented in Section 4.1. \\

Following these results, we inspected the \texttt{accelsearch} program to determine if the loss in sensitivity for broad pulses is solely due to the effect of red noise. This investigation was further motivated by the inconsistency between the radiometer predictions and the calculations from the injections under ideal, white noise conditions reported in \citet{lbh+15}. An issue in \texttt{accelsearch} was identified and corrected. \\ 

Despite the noticeable improvement in the PALFA sensitivity when using the FFA to look for long-period pulsar signals in survey data, both the Fourier-based and FFA-based search types are not able to recover weak signals that should be detectable according to the radiometer equation. Nevertheless, this analysis demonstrates the ability of a time-domain search technique to outperform a Fourier domain technique when applied in large-scale pulsar surveys and suggests that the PALFA survey should discover long-period pulsars via the new implementation of \texttt{ffaGo} in the data processing pipeline.  \\

\begin{table*}
  \setlength{\tabcolsep}{0.4cm}
  \caption{Parameters of the synthetic pulsar signals used in the FFA sensitivity analysis.}
  \label{tab:b}
  \begin{tabular}{p{8cm} c }
    \hline
    Parameter & Possible Values \\
    \hline \hline
    Period (ms) & 533.3, 1657.5, 3927.0, 5581.9, 10965.5, 14965.5*, 21427.7* \\
    DM (pc cm$^{-3}$) & 10, 40, 150, 600 \\
    FWHM ($\%$ Phase) & 0.5*, 1.5, 2.6, 5.9, 11.9 \\
    \hline
  \end{tabular}
  \begin{tablenotes} 
   \small
   \item * New trial elements that were not included in the sensitivity analysis conducted in \citet{lbh+15}. 
  \end{tablenotes}
\end{table*}

\section{Preliminary Results from the PALFA Survey}
Along with the addition of the FFA search in the PALFA \texttt{PRESTO}-based pipeline, we have modified some of the search parameters of the earlier version of the Fourier-based periodicity search. \\
  
The number of harmonics incoherently summed in the \textit{zero-acceleration} search, which is optimized to identify isolated pulsars, was initially set to 16. The new version of the search now sums up to 32 harmonics to increase our sensitivity to signals having narrow pulse profiles. This required changes to \texttt{accelsearch}. Doubling the number of harmonics summed in the \textit{zero-acceleration} search approximately triples the computation time for this specific Fourier-domain search. This remains a small fraction (less than 5$\%$) of the overall processing time. \\
  
Also, we lowered the limit on the lowest frequency of the highest harmonic to search. This parameter was previously set to 1 Hz and 2 Hz for accelerated and non-accelerated searches, respectively, and is now reduced to 0.5 Hz and 1 Hz, respectively. While this modification increases our sensitivity to pulsars having low-modulation frequencies, it potentially increases the resulting number of false positive candidates. \\


Following the recent implementation of the FFA and the above changes, the PALFA \texttt{PRESTO}-based pipeline has discovered five new sources\footnote{Discovery plots of those sources are available at \url{http://www.naic.edu/~palfa/newpulsars/}} with periods longer than a few hundreds of milliseconds: PSRs J1843+01 (P=1.267 sec), J1911+13 (P=0.300 sec), J1913+05 (P=0.662), J1914+08 (P=0.456 sec) and J1924+19 (1.278 sec). The FFA detected all five pulsars. Additionally, the Quicklook pipeline discovered J1901+11, a 409-ms pulsar which was later re-detected with both the FFA and the FFT searches of the full-resolution PALFA pipeline. Finally, \texttt{accelsearch} has found 3 pulsars with periods shorter than 100 ms, too short to be detected by \texttt{ffaGo}. \\

One pulsar, PSR J1913+05 (P=0.662 sec), was uniquely detected by the FFA search. It is interesting to note that it is not the slowest pulsar, but the weakest among the new discoveries. We estimate its flux density to be 11 $\mu$Jy with a pulse duty cycle $\delta$ of 2 $\%$ when using the FWHM as the pulse width. This demonstrates that in addition to be more sensitive to long period pulsars, \texttt{ffaGo} can outperform an FFT for weak pulsars with significant harmonic structures. This is consistent with our results from the simulations described in Section 5.2. \\

PSR J1911+13 has a rotation period of 0.300 sec, was found at a DM of 322.3 pc cm$^{-3}$ in inner Galaxy data, and appears to be a nulling pulsar. Both frequency-domain and time-domain analyses identified the source. Results from future timing observations of this nulling pulsar will be provided in a separate paper.  \\

Pulsars with spin periods shorter than 0.5 sec were detected by \texttt{ffaGo} via their second sub-harmonic, while it detected the ones with longer periods at their fundamental frequency. Timing solutions are not yet available for these new discoveries; precise parameters will be included in a future paper.\\

Thus far, our FFA pipeline detected more than 50 known sources having periods in the FFA range. There are a few instances where the FFA detected the second sub-harmonics of pulsars. All of these have periods shorter than 500 ms. Moreover, two known sources were re-detected by the FFA in beams a few arc-minutes away from the true pulsar positions, but were not detected by the FFT search. This is therefore promising and demonstrates once more the ability of \texttt{ffaGo} to detect pulsar signals in the survey.\\

\section{Conclusion}
In this paper, we have discussed the use of a FFA-based search, \texttt{ffaGo}, in the PALFA pulsar survey. In the PALFA implementation, \texttt{ffaGo} searches for periodic signals with 500 ms $\leqslant$ P $\leqslant$ 30 sec in time series dedispersed at DM values under 3265 pc cm$^{-3}$. \\

We compared the FFA program to \texttt{PRESTO}'s frequency-domain periodicity search, \texttt{accelsearch}, using a constructed dataset of synthetic pulsars having periods between 2 sec and 20 sec, with pulse duty cycles $\delta$ ranging from 0.5$\%$ and 20$\%$. Results showed that the FFA exceeds the performance of the FFT in the white noise regime in the case of long-period pulsars, especially when the signals have low harmonic content. We then selected a variety of long-period pulsar observations with periods between 1.32 sec and 4.6 sec discovered with PALFA and compared the response of both algorithms when searching in the presence of red noise and interference. The time-domain algorithm successfully detected all sources at their fundamental frequencies, at $\rm S/N_{\rm modified}$ values significantly larger than the detection threshold set in the pipeline. \\

The sensitivity of the PALFA survey was then assessed by conducting an analysis where we used ffaGo to recover a variety of synthetic pulsars injected into real PALFA survey observations. Comparing our results to those obtained by \citet{lbh+15}, we showed that for a pulse width of 2.6\%, the FFA outperforms the FFT for DM $\lesssim$ 40 pc cm$^{-3}$ and for periods as short as $\sim$500 ms, and the survey sensitivity is improved by at least a factor of two for periods $\gapp$ 6 sec. For the same width, the FFA exceeds the performance of the FFT for all trial DMs for periods longer than 5 sec. Moreover, for these periods, the sensitivity of the survey increases by at least a factor of three for pulsars having width $\gapp$ 11.9\% for all trial DMs. For periods greater than $\sim$ 8 sec, the FFA performs better than the FFT for all tested values of DM and all pulse FWHMs. This simulation demonstrated that the coherent summing of all harmonics greatly enhances the sensitivity of pulsars survey in the red noise regime. \\

As of now, our FFA search has uniquely discovered one pulsar and four others were discovered by both the FFA and the FFT searches of the PALFA survey. It has also re-detected more than 50 known pulsars that were present in the data. We are optimistic that our implementation of a FFA pipeline in the PALFA survey will lead to the discovery of new long-period pulsars in the future.\\ 

\section*{Acknowledgements}
We gratefully thank Vlad Kondratiev for useful discussions and his valuable assistance in the comparison of the different metrics used to evaluate FFA-generated profiles. EP acknowledges the support of NSERC (CGS M) and FQRNT B1. VMK acknowledges support from an NSERC Discovery grant and Herzberg Award, the Canada Research Chairs program, the Canadian Institute for Advanced Research, and FRQ-NT. SR is a Senior Fellow in the Canadian Institute for Advanced Research. PS is a Covington Fellow at DRAO. MAM is supported by NSF award number 1458952. WWZ is supported by the CAS Pioneer Hundred Talents Program. J.W.T.H. acknowledges funding from an NWO Vidi fellowship and from the European Research Council under the European Union's Seventh Framework Programme (FP/2007-2013) / ERC Starting Grant agreement nr. 337062 ("DRAGNET"). JSD was supported by the NASA Fermi program. KS, SR and FC are supported by the NANOGrav NSF Physics Frontiers Center award number 1430284. Pulsar research at UBC is supported by an NSERC Discovery Grant and by the Canadian Institute for Advanced Research. 

The Arecibo Observatory is operated by SRI International under a cooperative agreement with the National Science Foundation (AST-1100968), and in alliance with Ana G.M\'{e}ndez-Universidad Metropolitana, and the Universities Space Research Association. The CyberSKA project was funded by a CANARIE NEP-2 grant. 
Computations were made on the supercomputer Guillimin at McGill University, managed by Calcul Qu\'{e}bec and Compute Canada. The operation of this supercomputer is funded by the Canada Foundation for Innovation (CFI), NanoQu\'{e}bec, RMGA and the Fonds de recherche du Qu\'{e}bec$-$Nature et technologies (FRQ-NT).
\section*{Appendix}
\section*{Comparison between Metric A and Metric B}
The most important distinction between Metric A and Metric B is the calculation of a profile's standard deviation. For Metric A, this value is constant for a given trial period (i.e., the M folded profiles for a given trial period will have the same value of standard deviation), while it is profile-dependent for S/N calculations with Metric B. Analytically, regardless of the data distribution, a consequence of calculating the standard deviation directly on the profiles is that there is a greater spread in the standard deviation, especially significant in the case where fewer padded samples are folded. Excluding a 20$\%$ window around the profile peak in the calculations further decreases the number of bins within a profile, resulting in even larger standard deviation values compared to Metric A. However, when using real, non Gaussian-distributed data, removing some of the high-valued bins will reduce the sum of the squared deviations, therefore decreasing the standard deviation. It is therefore nontrivial to predict the expected reduction in S/N (or increase in standard deviation) from Metric B as M, the number of folded profiles, and z, the number of padded profiles, change. The simulations presented in Section 3.3 show that the S/N values from Metric B are the lowest for broad, very long-period pulsars. This suggests that the effective increase in the standard deviation in Metric B, which is more significant at longer periods (i.e., for smaller values of z), affects more the S/N calculations than the decrease in standard deviation resulting from the exclusion of the profile peak.

\section*{Profile evaluation with Metric C}
To evaluate FFA-generated profiles, we designed a third profile significance metric in \texttt{ffaGo}, Metric C. Similarly to Metric B, Metric C excludes a 20\% window centered on the peak of the profile when calculating the median intensity of the off-pulse, $I_{\rm med,off}$. This median value is therefore the same as for Metric B. However, the standard deviation of the off-pulse profile is a scaled-down version of the standard deviation computed with Metric A. The mathematical expression for the S/N calculated with this third metric is as follows:
\begin{equation}
  \textrm{S/N} = \frac{ I_{\rm max} - I_{\rm med,off} }{(\sigma\sqrt{\rm{X}})\, \sqrt{0.8 (M - z)}},
\end{equation} \\
where the factor $\sqrt{0.8}$ accounts for the exclusion of the on-pulse portion of the profile. \\

\begin{figure}[h!]
    \includegraphics[width=\columnwidth]{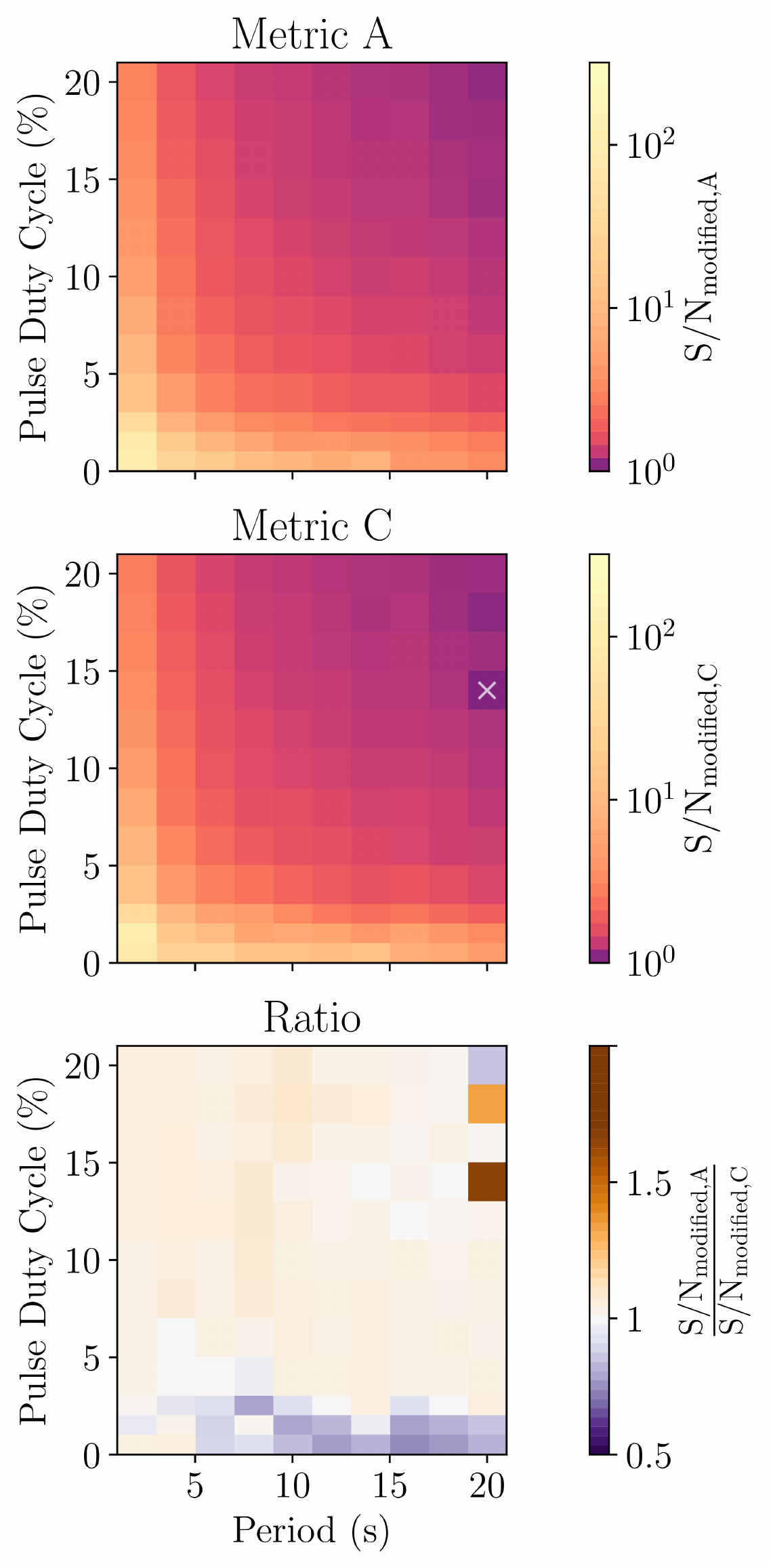}
    \caption{Response patterns of the two FFA significance metrics, Metric A (\textit{top panel}) and Metric C (\textit{middle panel}), obtained from the white noise simulation described in Section 3.3. The ratio of the two values of $\rm S/N_{\rm modified}$ is shown in the \textit{bottom panel}, which illustrates the relative responses of Metrics A and C. The values reported are the average $\rm S/N_{\rm modified}$ from the five simulations. The pixel with a white cross represents the trial having an average $\rm S/N_{\rm modified}$ below 6 (i.e., classified as a non-detection). Note that the scale for the top and the middle panels are logarithmic, while the bottom panel is displayed on a linear scale.}
    \label{fig:SNR_C}
\end{figure} 
We investigated the performance of Metric C when searching for pulsar signals in white noise, and compared it to Metric A. The same five datasets of 120 synthetic pulsars described in Section 3.3 were processed through \texttt{ffaGo} with Metric C, and we illustrate the results in Figure~\ref{fig:SNR_C}. This metric's response pattern is very similar to that of Metric A. Slightly larger values of $\rm S/N_{\rm modified}$ are obtained with Metric C when recovering narrow-pulsed pulsars, while $\rm S/N_{\rm modified}$ values are slightly lower when recovering broad, long-period signals. The mean value of the ratio matrix showed in the bottom panel of Figure~\ref{fig:SNR_C} is 1.01$\pm$0.10, which implies that there is no significant difference in the performance of the two metrics when considering the phase-space that our search is covering.\\

It is interesting to note that Metric C does not suffer the same sensitivity degradation in the long-period/large $\delta$ phase-space as Metric B. One can then conclude that it is not the exclusion of the on-pulse fraction that causes the most significant differences between Metric A and Metric B, but rather the calculation of the standard deviation directly on profiles rather than on the time series directly.\\

The performance of Metric C was also tested when searching for pulsars in the presence of red noise by performing two distinct analyses. First, we processed the selection of 12 real PALFA long-period pulsar observations (see Table~\ref{tab:long-p_pulsars}) through \texttt{ffaGo} while using Metric C. Metric A produced marginally larger values of $\rm S/N_{\rm modified}$ than Metric C for all pulsars. While Metric A identified the fundamental spin frequencies of all pulsars, there are two instances where Metric C identified the pulsar via its harmonics. \\

The second analysis consisted in the partial reproduction of the sensitivity analysis described in Section 5. Synthetic pulsars for a subset of the parameters listed in Table~\ref{tab:b} (only the trial DM = 150 pc cm$^{-3}$ was selected) were injected in one of the PALFA observation files used in the PALFA sensitivity analysis. The minimum mean flux density detectable by Metric C was then established. The results suggest that the performance of the two metrics are similar, regardless of the spin period of the pulsar. \\ 

We conclude that in presence of RFI and red noise, Metrics A and C behave similarly.
\bibliography{ffaarxiv.bib}

\begin{thebibliography}{}
\expandafter\ifx\csname natexlab\endcsname\relax\def\natexlab#1{#1}\fi
\providecommand{\url}[1]{\href{#1}{#1}}

\bibitem[{{Allen} {et~al.}(2013){Allen}, {Knispel}, {Cordes}, {Deneva},
  {Hessels}, {Anderson}, {Aulbert}, {Bock}, {Brazier}, {Chatterjee},
  {Demorest}, {Eggenstein}, {Fehrmann}, {Gotthelf}, {Hammer}, {Kaspi},
  {Kramer}, {Lyne}, {Machenschalk}, {McLaughlin}, {Messenger}, {Pletsch},
  {Ransom}, {Stairs}, {Stappers}, {Bhat}, {Bogdanov}, {Camilo}, {Champion},
  {Crawford}, {Desvignes}, {Freire}, {Heald}, {Jenet}, {Lazarus}, {Lee}, {van
  Leeuwen}, {Lynch}, {Papa}, {Prix}, {Rosen}, {Scholz}, {Siemens}, {Stovall},
  {Venkataraman}, \& {Zhu}}]{akc+13}
{Allen}, B., {Knispel}, B., {Cordes}, J.~M., {et~al.} 2013, \apj, 773, 91

\bibitem[{{Bhat} {et~al.}(2004){Bhat}, {Cordes}, {Camilo}, {Nice}, \&
  {Lorimer}}]{bccnl04}
{Bhat}, N.~D.~R., {Cordes}, J.~M., {Camilo}, F., {Nice}, D.~J., \& {Lorimer},
  D.~R. 2004, \apj, 605, 759

\bibitem[{{Cameron} {et~al.}(2017){Cameron}, {Barr}, {Champion}, {Kramer}, \&
  {Zhu}}]{cbc+17}
{Cameron}, A.~D., {Barr}, E.~D., {Champion}, D.~J., {Kramer}, M., \& {Zhu},
  W.~W. 2017, \mnras, 468, 1994

\bibitem[{{Chen} \& {Ruderman}(1993)}]{cr93}
{Chen}, K., \& {Ruderman}, M. 1993, \apj, 402, 264

\bibitem[{{Cordes} \& {Lazio}(2002)}]{cl02}
{Cordes}, J.~M., \& {Lazio}, T.~J.~W. 2002, ArXiv Astrophysics e-prints,
  astro-ph/0207156

\bibitem[{{Cordes} \& {McLaughlin}(2003)}]{cm03}
{Cordes}, J.~M., \& {McLaughlin}, M.~A. 2003, \apj, 596, 1142

\bibitem[{{Craft} {et~al.}(1968){Craft}, {Sutton}, \& {Comella}}]{csc68}
{Craft}, H.~D., {Sutton}, J.~M., \& {Comella}, J.~M. 1968, \nat, 219, 1237

\bibitem[{{Crawford} {et~al.}(2009){Crawford}, {Lorimer}, {Devour}, {Takacs},
  \& {Kondratiev}}]{cldtk09}
{Crawford}, F., {Lorimer}, D.~R., {Devour}, B.~M., {Takacs}, B.~P., \&
  {Kondratiev}, V.~I. 2009, \apj, 696, 574

\bibitem[{{Dewey} {et~al.}(1985){Dewey}, {Taylor}, {Weisberg}, \&
  {Stokes}}]{dtws85}
{Dewey}, R.~J., {Taylor}, J.~H., {Weisberg}, J.~M., \& {Stokes}, G.~H. 1985,
  \apjl, 294, L25

\bibitem[{{Eatough}(2007)}]{e07}
{Eatough}, R. 2007, in From Planets to Dark Energy: the Modern Radio Universe,
  92

\bibitem[{{Faulkner} {et~al.}(2004){Faulkner}, {Stairs}, {Kramer}, {Lyne},
  {Hobbs}, {Possenti}, {Lorimer}, {Manchester}, {McLaughlin}, {D'Amico},
  {Camilo}, \& {Burgay}}]{fsk+04}
{Faulkner}, A.~J., {Stairs}, I.~H., {Kramer}, M., {et~al.} 2004, \mnras, 355,
  147

\bibitem[{{Hibschman} \& {Arons}(2001)}]{ha01}
{Hibschman}, J.~A., \& {Arons}, J. 2001, \apj, 554, 624

\bibitem[{{Kaspi} \& {Beloborodov}(2017)}]{kb17}
{Kaspi}, V.~M., \& {Beloborodov}, A.~M. 2017, \araa, 55, 261

\bibitem[{{Kiddle} {et~al.}(2011){Kiddle}, {Andrecut}, {Brazier}, {Chatterjee},
  {Chen}, {Cordes}, {Curry}, {Este}, {Eymere}, {Federl}, {Fong}, {Grimstrup},
  {Guram}, {Kaspi}, {Klodzinski}, {Lazarus}, {Mahadevan}, {Mourad}, {Mourad},
  {Pragides}, {Rosolowsky}, {Said}, {Samoilov}, {Smith}, {Stairs}, {Tan},
  {Tan}, {Taylor}, \& {Willis}}]{kab+11}
{Kiddle}, C., {Andrecut}, M., {Brazier}, A., {et~al.} 2011, in Astronomical
  Society of the Pacific Conference Series, Vol. 442, Astronomical Data
  Analysis Software and Systems XX, ed. I.~N. {Evans}, A.~{Accomazzi}, D.~J.
  {Mink}, \& A.~H. {Rots}, 669

\bibitem[{{Kondratiev} {et~al.}(2009){Kondratiev}, {McLaughlin}, {Lorimer},
  {Burgay}, {Possenti}, {Turolla}, {Popov}, \& {Zane}}]{kml+09}
{Kondratiev}, V.~I., {McLaughlin}, M.~A., {Lorimer}, D.~R., {et~al.} 2009,
  \apj, 702, 692

\bibitem[{{Lazarus} {et~al.}(2015){Lazarus}, {Brazier}, {Hessels},
  {Karako-Argaman}, {Kaspi}, {Lynch}, {Madsen}, {Patel}, {Ransom}, {Scholz},
  {Swiggum}, {Zhu}, {Allen}, {Bogdanov}, {Camilo}, {Cardoso}, {Chatterjee},
  {Cordes}, {Crawford}, {Deneva}, {Ferdman}, {Freire}, {Jenet}, {Knispel},
  {Lee}, {van Leeuwen}, {Lorimer}, {Lyne}, {McLaughlin}, {Siemens}, {Spitler},
  {Stairs}, {Stovall}, \& {Venkataraman}}]{lbh+15}
{Lazarus}, P., {Brazier}, A., {Hessels}, J.~W.~T., {et~al.} 2015, \apj, 812, 81

\bibitem[{{Lipunov} {et~al.}(2005){Lipunov}, {Bogomazov}, \&
  {Abubekerov}}]{lba05}
{Lipunov}, V.~M., {Bogomazov}, A.~I., \& {Abubekerov}, M.~K. 2005, \mnras, 359,
  1517

\bibitem[{{Lorimer} \& {Kramer}(2004)}]{lk04}
{Lorimer}, D.~R., \& {Kramer}, M. 2004, {Handbook of Pulsar Astronomy}

\bibitem[{{Lorimer} {et~al.}(2006){Lorimer}, {Faulkner}, {Lyne}, {Manchester},
  {Kramer}, {McLaughlin}, {Hobbs}, {Possenti}, {Stairs}, {Camilo}, {Burgay},
  {D'Amico}, {Corongiu}, \& {Crawford}}]{lfl+06}
{Lorimer}, D.~R., {Faulkner}, A.~J., {Lyne}, A.~G., {et~al.} 2006, \mnras, 372,
  777

\bibitem[{{Lovelace} {et~al.}(1969){Lovelace}, {Sutton}, \& {Salpeter}}]{ls69}
{Lovelace}, R.~V.~E., {Sutton}, J.~M., \& {Salpeter}, E.~E. 1969, \nat, 222,
  231

\bibitem[{{Lyne} {et~al.}(2017){Lyne}, {Stappers}, {Bogdanov}, {Ferdman},
  {Freire}, {Kaspi}, {Knispel}, {Lynch}, {Allen}, {Brazier}, {Camilo},
  {Cardoso}, {Chatterjee}, {Cordes}, {Crawford}, {Deneva}, {Hessels}, {Jenet},
  {Lazarus}, {van Leeuwen}, {Lorimer}, {Madsen}, {McKee}, {McLaughlin},
  {Parent}, {Patel}, {Ransom}, {Scholz}, {Seymour}, {Siemens}, {Spitler},
  {Stairs}, {Stovall}, {Swiggum}, {Wharton}, {Zhu}, {Aulbert}, {Bock},
  {Eggenstein}, {Fehrmann}, \& {Machenschalk}}]{lsb+17}
{Lyne}, A.~G., {Stappers}, B.~W., {Bogdanov}, S., {et~al.} 2017, \apj, 834, 137

\bibitem[{Patel(2016)}]{Patel:Thesis:2016}
Patel, C. 2016, Master's thesis, McGill University, Montreal, QC, Canada

\bibitem[{{Petigura} {et~al.}(2013){Petigura}, {Marcy}, \& {Howard}}]{pmh13}
{Petigura}, E.~A., {Marcy}, G.~W., \& {Howard}, A.~W. 2013, \apj, 770, 69

\bibitem[{{Pfahl} {et~al.}(2005){Pfahl}, {Podsiadlowski}, \&
  {Rappaport}}]{ppr05}
{Pfahl}, E., {Podsiadlowski}, P., \& {Rappaport}, S. 2005, \apj, 628, 343

\bibitem[{{Ransom}(2001)}]{r01}
{Ransom}, S.~M. 2001, PhD thesis, Harvard University

\bibitem[{{Staelin}(1969)}]{s69}
{Staelin}, D.~H. 1969, IEEE Proceedings, 57, 724

\bibitem[{{Stovall} {et~al.}(2013){Stovall}, {Jenet}, {Siemens}, {Kaplan},
  {Creighton}, {Miller}, {Rodriguez-Zermeno}, {Banaszak}, {Biwer}, {Ceballos},
  {Cohen}, {Day}, {Ford}, {Flanigan}, {Garcia}, {Hinojosa}, {Leake},
  {Martinez}, {Mata}, {Miller}, {Murray}, {Rivera}, {Reser}, {Rohr}, {Rudnik},
  {Walker}, {Wells}, {Consortium}, {Consortium}, {Drift Consortium}, \&
  {Consortium}}]{sjsk+13}
{Stovall}, K., {Jenet}, F., {Siemens}, X., {et~al.} 2013, in American
  Astronomical Society Meeting Abstracts, Vol. 221, American Astronomical
  Society Meeting Abstracts \#221, 154.05

\bibitem[{{Swiggum} {et~al.}(2014){Swiggum}, {Lorimer}, {McLaughlin}, {Bates},
  {Champion}, {Ransom}, {Lazarus}, {Brazier}, {Hessels}, {Nice}, {Ellis},
  {Senty}, {Allen}, {Bhat}, {Bogdanov}, {Camilo}, {Chatterjee}, {Cordes},
  {Crawford}, {Deneva}, {Freire}, {Jenet}, {Karako-Argaman}, {Kaspi},
  {Knispel}, {Lee}, {van Leeuwen}, {Lynch}, {Lyne}, {Scholz}, {Siemens},
  {Stairs}, {Stappers}, {Stovall}, {Venkataraman}, \& {Zhu}}]{slm14}
{Swiggum}, J.~K., {Lorimer}, D.~R., {McLaughlin}, M.~A., {et~al.} 2014, \apj,
  787, 137

\bibitem[{von Mises(1918)}]{von18}
von Mises, R. 1918, Phys. Z., 19, 490.
\newblock \url{http://ci.nii.ac.jp/naid/10006156040/en/}

\bibitem[{{Young} {et~al.}(1999){Young}, {Manchester}, \& {Johnston}}]{ymj99}
{Young}, M.~D., {Manchester}, R.~N., \& {Johnston}, S. 1999, \nat, 400, 848

\bibitem[{{Zhang} {et~al.}(2000){Zhang}, {Harding}, \& {Muslimov}}]{zhm00}
{Zhang}, B., {Harding}, A.~K., \& {Muslimov}, A.~G. 2000, in Bulletin of the
  American Astronomical Society, Vol.~32, American Astronomical Society Meeting
  Abstracts \#195, 880

\bibitem[{{Zhu} {et~al.}(2014){Zhu}, {Berndsen}, {Madsen}, {Tan}, {Stairs},
  {Brazier}, {Lazarus}, {Lynch}, {Scholz}, {Stovall}, {Ransom}, {Banaszak},
  {Biwer}, {Cohen}, {Dartez}, {Flanigan}, {Lunsford}, {Martinez}, {Mata},
  {Rohr}, {Walker}, {Allen}, {Bhat}, {Bogdanov}, {Camilo}, {Chatterjee},
  {Cordes}, {Crawford}, {Deneva}, {Desvignes}, {Ferdman}, {Freire}, {Hessels},
  {Jenet}, {Kaplan}, {Kaspi}, {Knispel}, {Lee}, {van Leeuwen}, {Lyne},
  {McLaughlin}, {Siemens}, {Spitler}, \& {Venkataraman}}]{zbm+14}
{Zhu}, W.~W., {Berndsen}, A., {Madsen}, E.~C., {et~al.} 2014, \apj, 781, 117

\end{thebibliography}
\bibliographystyle{aasjournal}
\end{document}